\documentclass[a4paper, amsfonts, amssymb, amsmath, reprint, showkeys, footinbib, twoside,superscriptaddress,floatfix,longbibliography]{revtex4-1}

\usepackage{graphicx}%
\usepackage{dcolumn}%
\usepackage{bm}%
\usepackage{float}
\usepackage{xcolor}
\usepackage{mhchem}
\usepackage{gensymb}
\usepackage[slantedGreek]{newtxmath}
\usepackage{comment}
\usepackage[pdftex, bookmarks, pdffitwindow=false, pdfstartview=FitH, pdfdisplaydoctitle, colorlinks, plainpages=false, pdftitle={},pdfauthor={}, pdfpagelabels, hypertexnames, citecolor={blue!50!black},linkcolor={blue!50!black}, urlcolor={blue!50!black}, pdflang={en}, hyperfootnotes=false, breaklinks]{hyperref}
\usepackage{siunitx}

\newcommand{\figLabelCapt}[1]{(\MakeLowercase{{#1}})}
\newcommand{\refSub}[2]{\hyperref[#2]{\ref{#2}\figLabelCapt{#1}}}
\newcommand{\figref}[1]{Fig.~\ref{#1}}
\newcommand{\figrefsub}[2]{Fig.~\refSub{#2}{#1}}
\newcommand{\figsrefsub}[2]{Figs.~\refSub{#2}{#1}}

\makeatletter
\renewcommand{\@seccntformat}[1]{}
\makeatother

\makeatletter
\def\@bibdataout@aps{
 \immediate\write\@bibdataout{
 @CONTROL{
   apsrev41Control, author="48",editor="1",pages="0",title="0",year="1"
 }}
 \if@filesw
  \immediate\write\@auxout{\string\citation{apsrev41Control}}
 \fi
}
\makeatother

\begin{document}
%\preprint{}

\title{Hydrogen–helium immiscibility boundary in gas-giant planetary interiors from machine-learning molecular dynamics} 

\author{Xiaoyu Wang}
\affiliation{Department of Chemistry, UC Berkeley, California 94720, United States}

\author{Sebastien Hamel}
%\email{hamel2@llnl.gov}
\affiliation{Lawrence Livermore National Laboratory, Livermore, CA, USA}

\author{Bingqing Cheng}
\email{bingqingcheng@berkeley.edu}
\affiliation{Department of Chemistry, UC Berkeley, California 94720, United States}
\affiliation{The Institute of Science and Technology Austria, Am Campus 1, 3400 Klosterneuburg, Austria}
\affiliation{Materials Sciences Division, Lawrence Berkeley National Laboratory, Berkeley, CA, USA}
\affiliation{Chemical Sciences Division, Lawrence Berkeley National Laboratory, Berkeley, CA, USA}
\affiliation{Bakar Institute of Digital Materials for the Planet, UC Berkeley, California 94720, United States}

\date{\today}%

\begin{abstract}
The location of the hydrogen–helium (H/He) immiscibility boundary controls whether and where helium rain occurs in giant planets, yet it remains uncertain because high-pressure experiments are challenging and \emph{ab initio} simulations are limited in system size and simulation time. 
We map this boundary by computing composition-dependent chemical potentials from large-scale molecular dynamics driven by machine learning potentials trained on three density functional approximations (PBE, vdW-DF, and the hybrid HSE). 
Across the pressure range of 100–1000~GPa, the demixing temperatures are typically $\sim$2000~K lower than previous \emph{ab initio} simulations using small system sizes, and this discrepancy is larger than the estimated error from the chosen density functionals, machine learning potentials, neglect of nuclear quantum effects, and statistical uncertainties.
Fitting the H/He mixing free energy to a Redlich–Kister regular solution model rationalizes the thermodynamic driving force for phase separation, and provides a predictive representation of the boundary. 
Comparing with current planetary interior profiles indicates that helium rain is plausible in Saturn but unlikely in the warmer interior of Jupiter.
Our results narrow the uncertainty in the H/He immiscibility boundary and provide inputs for planetary models that couple demixing, heat transport, and composition gradients in gas giants.
\end{abstract}

\maketitle

\section{Introduction}
Gas giants such as Jupiter, Saturn, and Jovian extrasolar planets are composed predominantly of hydrogen and helium~\cite{helled2020understanding}.
In the deep interiors of these planets at hundreds of gigapascals and thousands of kelvins, 
helium is hypothesized to become immiscible in hydrogen and to precipitate as helium-rich droplets that sink toward the planetary interior under gravity. This phase separation and subsequent gravitational settling of helium-rich components are commonly referred to as ``helium rain''~\cite{stevenson1977phase, stevenson1977dynamics, Helled2019interiors, mcmahon2012properties}.
We use ``helium rain'' in the conventional astrophysical sense, to denote the precipitation and gravitational settling of helium-rich droplets, not a cyclic meteorological process.
This mechanism can help reconcile a spectrum of observations from Voyager~\cite{campbell1985gravity}, Pioneer~\cite{campbell1985gravity}, Juno~\cite{bolton2017jupiter}, and Cassini~\cite{spilker2019cassini} missions:
The rainout of He-rich droplets from outer envelopes would account for the atmospheric helium depletion~\cite{young2003galileo} in Jupiter~\cite{von1998helium} and Saturn~\cite{fortney2023saturn}, relative to the protosolar helium atomic fraction of $x_{\rm He} = 0.089$~\cite{Proffitt1994proto, bahcall1995solar}. 
The gravitational energy released by falling droplets provides an additional heat source that helps explain the excess luminosity of Jupiter~\cite{low1966observations} and Saturn~\cite{stevenson1977dynamics}.
Moreover, the resulting compositional stratification of He-rich mixtures can create distinct layered structures within the planets, altering their internal structures, dynamics, and magnetic fields~\cite{militzer2016understanding, Helled2019interiors, militzer2024study, duarte2018physical, howard2024evolution}. 

While the hypothesis of helium rain is incorporated into the current planetary models for Jupiter and Saturn~\cite{militzer2024study,howard2024evolution}, all these models remain speculative due to the large uncertainty about the H/He immiscibility boundary.
For example, several Jupiter models suggest the presence of 3 to 6 distinct interior layers, all of which are consistent with the gravity field data of Juno, despite varying structural assumptions~\cite{militzer2024study}. 
The extreme planetary $P$–$T$ conditions pose tremendous experimental challenges, leading to limited data along the immiscibility boundary: Only one high-pressure experimental study on H/He demixing with laser-shock~\cite{brygoo2021evidence} is currently available, which reports a demixing temperature of 10,200~K at 150~GPa for a H/He mixture of $x_{\rm He} = 0.11$. 
Theoretical calculations based on quantum mechanics~\cite{lorenzen2009demixing, morales2009, lorenzen2011metallization, morales2013hydrogen, schottler2018ab, chang2024theoretical} have thus played an important role in determining the immiscibility boundary:
Most \emph{ab initio} molecular dynamics (AIMD) studies based on density functional theory (DFT) predict that H/He demixing occurs at 5000–6000~K under 150~GPa~\cite{morales2013hydrogen, schottler2018ab} for $x_{\rm He}\approx0.09$. 
There are several challenges in the \emph{ab initio} modeling of H/He demixing.
First, DFT relies on approximations of the exchange-correlation (XC) functionals which can cause systematic errors in describing atomic interactions~\cite{mori2008localization, cohen2012challenges, burke2012perspective, sim2022improving, cozza2026denser}, and the effect of such approximations on the H/He immiscibility remains unclear~\cite{clay2016benchmarking, schottler2018ab, helled2020understanding, chang2024theoretical}. 
Second, estimating the non-ideal entropy of mixing requires computationally demanding free energy calculations such as thermodynamic integration (TI)~\cite{morales2009, schottler2018ab}. 
Third, DFT studies are often constrained to system sizes of a few hundred atoms and simulation durations of a few picoseconds due to the computational expense.
The finite size effects can be significant, especially for phase transition phenomena~\cite{cheng2020evidence, cheng2021reply}.
Recently, machine learning potentials (MLPs) have been introduced to overcome this obstacle by first learning atomic interactions from DFT energies and forces of training configurations~\cite{Deringer2019}, and subsequently driving large-scale MD simulations. 
A recent study using MLPs reports much higher demixing temperatures up to 8500~K at 150~GPa~\cite{chang2024theoretical}. 

Here, we perform large-scale MD simulations driven by three sets of MLPs that are trained based on three different DFT functionals: the Perdew–Burke–Ernzerhof (PBE)~\cite{perdew1996generalized} and van der Waals density functional (vdW-DF)~\cite{lee2010higher} that are both widely used for such high-pressure mixtures, and the Heyd–Scuseria–Ernzerhof (HSE) hybrid functional~\cite{heyd2003hybrid, heyd2006hybrid} for its inclusion of exact exchange and superior performance in benchmark studies~\cite{clay2016benchmarking}.
We benchmark the MLPs based on energetic, structural, and thermodynamic properties, as detailed in Supplementary Table~S1 and Supplementary Figs.~S3–S8.
We explore a broad range of $P$–$T$ conditions ($P$: 100–1000~GPa; $T$: 1000–12000~K) that span the typical interior conditions of gas giants. 
We accurately account for the thermal fluctuations with entropic effects by computing the chemical potentials ($\mu$) at different helium concentrations using the S0 method~\cite{cheng2022computing}.
We then use a Redlich–Kister regular solution model~\cite{redlich1948thermodynamics} to rationalize the thermodynamic driving force behind H/He phase separation, and to help understand the discrepancies with previous AIMD studies~\cite{lorenzen2011metallization,morales2009, morales2013hydrogen, schottler2018ab}.
Finally, we discuss the implications of our findings for the evolution of Jupiter and Saturn by juxtaposing our results with current planetary models~\cite{militzer2013ab, nettelmann2008ab, schottler2018ab, nettelmann2013saturn, howard2024evolution, mankovich2020evidence}.

\section{Results}
\subsection{Machine learning potentials for H/He mixtures}
We constructed MLPs for simulating the phase behaviors of H/He mixtures under planetary conditions ($10^2$–$10^3$~GPa and $10^3$–$10^4$~K).
Separate MLPs were fitted to data generated with PBE, vdW-DF, and HSE, in order to understand the impact of the underlying XC approximations. 
The vdW-DF training set comprises 13,389 configurations with varying helium fractions across a broad range of $P$–$T$ conditions (1–1200~GPa and 500–16000~K), which was generated via a combination of strategies: AIMD simulations, random structure searches, active learning, and adapting from previous AIMD calculations for high-pressure H/He mixtures~\cite{schottler2018ab}. 
We obtained the training set for the PBE MLP by recomputing the PBE energies and forces for all these configurations, as well as supplementing with a previous pure hydrogen training set~\cite{cheng2020evidence}. 
To make the DFT computational cost of generating the training set for the HSE MLP tractable, 
we employed a $\Delta$-learning strategy~\cite{singraber2019parallel} that treats the vdW-DF MLP as a baseline and learns the difference between vdW-DF and HSE energy surfaces ($\Delta$(vdW-DF$\rightarrow$HSE)). 
Specifically, we selected $\sim$2000 configurations from the full training set, recomputed their HSE energies and forces, and fitted the $\Delta$-learning HSE MLP.
The production HSE MLP adds the $\Delta$-learning MLP $\Delta$(vdW-DF$\rightarrow$HSE) to the baseline vdW-DF MLP.
As a validation of this $\Delta$-learning strategy, we built an analogous $\Delta$(PBE$\rightarrow$vdW-DF) potential using the PBE MLP as a baseline, and compared it to the direct-learning vdW-DF MLP. 

There are many MLP architectures available~\cite{wang2025design}, with different trade-offs between computational efficiency and predictive accuracy.
For the vdW-DF dataset, we tested three MLP architectures: the parallelizable and CPU-efficient N2P2~\cite{singraber2019parallel}, the more recent and accurate MACE~\cite{batatia2022mace} and CACE~\cite{cheng2024cartesian} models, as detailed in Methods.
The performance of each fitted MLP was validated by comparing its predicted equations of state (EOSs) and radial distribution functions (RDFs) against reference DFT data over a wide range of pressures (100–1200~GPa) and temperatures (2000–10000~K).
Even though N2P2 gives approximately twice the energy and force errors as MACE and CACE, the three vdW-DF MLPs trained on these architectures, as well as the aforementioned $\Delta$(PBE$\rightarrow$vdW-DF) MLP, yield EOS and RDF results that are consistent with vdW-DF AIMD results, as detailed in the Supplementary Figs.~S3–S8.
As an illustration, the pressure–density ($P$–$\rho$) relationship is presented for five H/He mixtures ($x_{\rm He}$ = 0, 0.25, 0.5, 0.75, and 1.0, color coded) at $T$ = 2000~K (\figrefsub{Fig:S0_mu_mixG}{a}), 6000~K (\figrefsub{Fig:S0_mu_mixG}{b}), and 10000~K (\figrefsub{Fig:S0_mu_mixG}{c}), and these four MLPs produce nearly identical pressure values matching the vdW-DF AIMD data~\cite{schottler2018ab}.
These results confirm that all the MLPs are sufficiently accurate for these thermodynamic properties.

\subsection{Chemical potentials of H and He in the mixtures}
\begin{figure*}[htbp!]
    \centering
   \includegraphics[width=\textwidth]{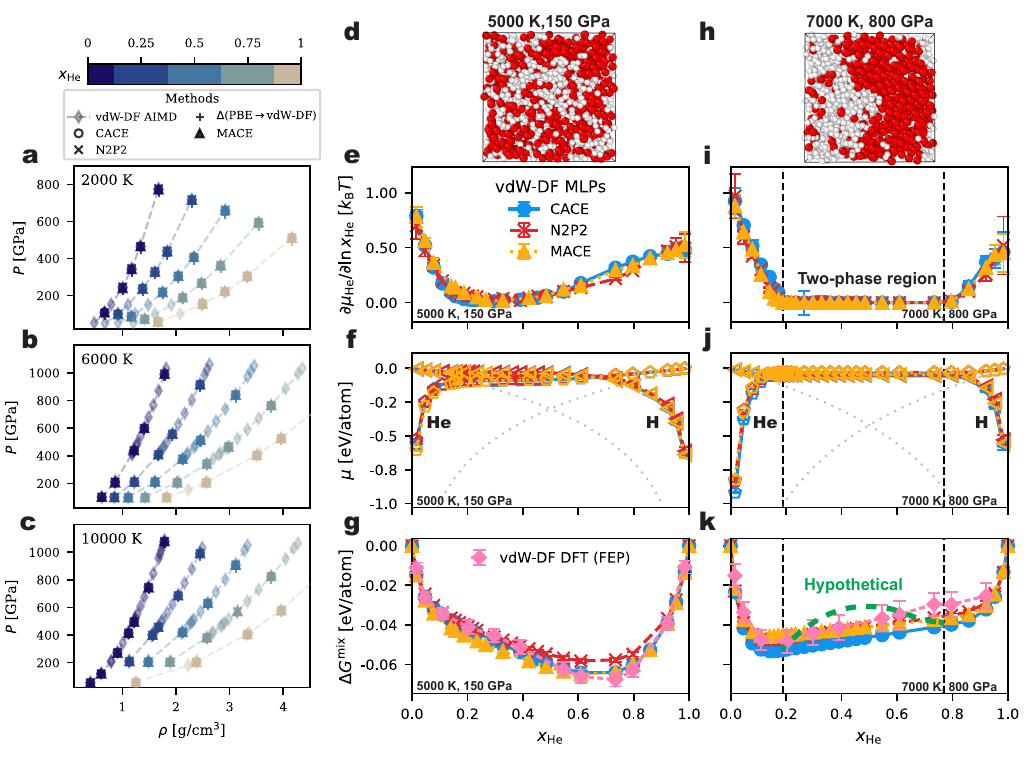}
    \caption{
    Thermodynamic behaviors of H/He mixtures predicted by different machine learning potentials (MLPs).
    \figLabelCapt{a–c} Pressure ($P$) as a function of density ($\rho$) for five H/He mixtures (helium atomic fraction $x_{\rm He}$ = 0, 0.25, 0.5, 0.75, and 1.0) at \figLabelCapt{a} 2000~K, \figLabelCapt{b} 6000~K, and \figLabelCapt{c} 10000~K. Results are from different vdW-DF MLPs: CACE (circles), MACE (triangles), N2P2 (crosses), and a $\Delta$-learning vdW-DF model (pluses) built upon the PBE MLP baseline. All results are compared to reference vdW-DF data (dashed lines with diamonds) from AIMD calculations~\cite{schottler2018ab}.
    \figLabelCapt{d, h} Snapshots from MLP MD simulations showing: \figLabelCapt{d} homogeneous mixture and \figLabelCapt{h} phase separation. White and red particles indicate hydrogen and helium atoms, respectively.
    \figLabelCapt{e, i} Derivative of the helium chemical potential $({\partial \mu_{\rm He}}/{\partial \ln x_{\rm He}})_{T, P}$ computed from the S0 method~\cite{cheng2022computing} with error bars indicating $\pm1$ standard error propagated from the linear-fit covariance matrices for $S^0$. The MD simulations were performed using MLPs trained on vdW-DF. Blue, red, and yellow symbols indicate MLP architectures employing CACE, N2P2, and MACE, respectively.
    The vertical dashed black lines represent the equilibrium $x_{\rm He}$ of the He-poor ($x_{\rm He} = 0.19$) and He-rich ($x_{\rm He} = 0.77$) phases during demixing at 7000~K and 800~GPa. 
    \figLabelCapt{f, j} Computed chemical potentials of H (triangles) and He (pentagons), compared with ideal solution behaviors (dotted light gray lines). 
    Error bars indicate uncertainties propagated from $({\partial \mu_{\rm He}}/{\partial \ln x_{\rm He}})_{T, P}$ through the numerical integration.
    \figLabelCapt{g, k} Mixing free energy per atom for H/He mixtures ($\Delta G^{\rm mix}$) as a function of helium fractions. Pink diamonds denote the $\Delta G^{\rm mix}$ at the vdW-DF DFT level, promoted from the vdW-DF N2P2 MLP to DFT using free-energy perturbation (FEP), with error bars indicating $\pm1$ standard error of the mean across 100 uncorrelated configurations at each $x_{\rm He}$.
    The schematic dashed green line in \figLabelCapt{k} represents the hypothetical $\Delta G^{\rm mix}$ curve for a homogeneous H/He mixture inside the demixing region. 
}
    \label{Fig:S0_mu_mixG}
\end{figure*}

We applied the direct-learning vdW-DF MLPs based on N2P2, MACE, and CACE in large-scale MD simulations, and checked the influence of MLP architecture on predicting H/He demixing behaviors.
Two sets of thermodynamic conditions representative of planetary conditions,
$T=5000$~K, $P=150$~GPa, and $T=7000$~K, $P=800$~GPa, were used.
The H/He mixture shows distinct behaviors:
As illustrated in \figsrefsub{Fig:S0_mu_mixG}{d, h},
the system is fully mixed at $T=5000$~K, $P=150$~GPa, and shows phase separation at $T=7000$~K, $P=800$~GPa.

To compute the chemical potentials of the H/He mixture,
we employed the S0 method~\cite{cheng2022computing},
which circumvents the convergence issue of traditional particle insertion methods~\cite{frenkel2023understanding}.
At certain $P$–$T$ and for a given composition, the derivative of the chemical potential of helium ($\mu_{\rm He}$) with respect to composition $\left(\partial \mu_{\rm He}/\partial \ln x_{\rm He}\right)_{T, P}$ is given by the S0 method: 
\begin{equation}
\left(\frac{\partial \mu_{\rm He}}{\partial \ln x_{\rm He}}\right)_{T, P}= \frac{k_{\rm B} T}{x_{\rm H} S_{\text{He–He}}^0 + x_{\rm He} S_{\text{H–H}}^0 - 2 \sqrt{x_{\rm H} x_{\rm He} }S_{\text{H–He}}^0},
\label{Eq: mu_i_liq_state}
\end{equation}
where $S_{\text{He–He}}^0$, $S_{\text{H–He}}^0$, and $S_{\text{H–H}}^0$ are the static structure factors between different pairs of atoms in the limit of infinite wavelength~\cite{kirkwood1951statistical, ben2006molecular}.

At $T=5000$~K, $P=150$~GPa, and $T=7000$~K, $P=800$~GPa, we ran equilibrium NPT MD simulations using a system size of 3456 atoms, driven by the three vdW-DF MLPs with different architectures. 
We sampled a series of compositions with the helium atomic fraction, $x_{\rm He}$, ranging from 0.015 to 0.98 in 26 intervals.
The computed chemical potential derivatives $\left(\partial \mu_{\rm He}/\partial \ln x_{\rm He}\right)_{T, P}$ are shown in \figsrefsub{Fig:S0_mu_mixG}{e, i}.
The N2P2, MACE, and CACE results agree well for both conditions.
Notably, the computed $\left(\partial \mu_{\rm He}/\partial \ln x_{\rm He}\right)_{T, P}$ at $T=7000$~K, $P=800$~GPa (\figrefsub{Fig:S0_mu_mixG}{i}) exhibits a near-zero plateau between $x_{\rm He\text{-}poor} = 0.19 $ and $x_{\rm He\text{-}rich} = 0.77$ (dashed black lines), signaling two-phase coexistence, consistent with the microscopic structural configuration observed in \figrefsub{Fig:S0_mu_mixG}{h}. 
The chemical potentials of helium and hydrogen are shown in \figsrefsub{Fig:S0_mu_mixG}{f, j}, obtained by integrating these derivatives $\left(\partial \mu_{\rm He}/\partial \ln x_{\rm He}\right)_{T, P}$ over composition. 
In the two-phase region at $T=7000$~K, $P=800$~GPa, $\mu_{\rm He}$ and $\mu_{\rm H}$ are constant, and only the proportions of the He-rich and He-poor phases change.

The per-atom Gibbs free energy of mixing ($\Delta G^{\rm mix}$) is:
\begin{equation}
\Delta G^{\rm {mix}} \left(x_{\rm He}\right)=x_{\rm He} \mu_{\rm He} (x_{\rm He}) + \left(1-x_{\rm He}\right) \mu_{\rm H} (x_{\rm He}).
\end{equation}
At 5000~K and 150~GPa, the $\Delta G^{\rm mix}$ curve in \figrefsub{Fig:S0_mu_mixG}{g} is fully convex, indicating complete miscibility.
In contrast, at 800~GPa and 7000~K, the mixing curve is linear in the demixing region ($x_{\rm He} = 0.19\text{–}0.77$) that corresponds to $\left(\partial \mu_{\rm He}/\partial \ln x_{\rm He}\right)_{T, P} = 0$ in \figrefsub{Fig:S0_mu_mixG}{i}.  
A hypothetical homogeneous mixture there would be thermodynamically unstable, as its $\Delta G^{\rm mix}$ (schematic dashed green line in~\figrefsub{Fig:S0_mu_mixG}{k}) exceeds that of the equilibrium phase-separated state (colored symbols). 
The phase separation also turns out to be kinetically rapid in our MD simulations: the separation completes within tens of picoseconds (\figrefsub{Fig:S0_mu_mixG}{h}), implying a low activation barrier for H/He demixing.

To quantify whether the use of the MLP introduces bias in the predicted mixing free energies, we further promoted the MLP-derived $\Delta G^{\rm mix}$ to the vdW-DF DFT level using free-energy perturbation (FEP). As shown by pink diamond symbols in \figsrefsub{Fig:S0_mu_mixG}{g, k}, the $\Delta G^{\rm mix}$ values at the DFT and at the MLP level show good agreement.
In the Supplementary Fig.~S15, we further show the agreement between MLP and DFT $\Delta G^{\rm mix}$ at two other conditions.
This suggests that the MLPs are able to accurately capture the thermodynamics of H/He mixing and demixing.

Taken together, all three vdW-DF MLPs produce consistent $\Delta G^{\rm mix}$ results for H/He mixtures and hence consistent predictions of phase separation.
Given the higher computational efficiency of the N2P2 MLP (see the Supplementary Information Section~IV), in what follows, we use it to sample the large $P$–$T$–$x_{\rm He}$ parameter space relevant to gas giants via high-throughput MD simulations. 

\subsection{Phase separation of H/He mixtures}

\begin{figure*}[htbp!]
    \centering
   \includegraphics[width=0.8\textwidth]{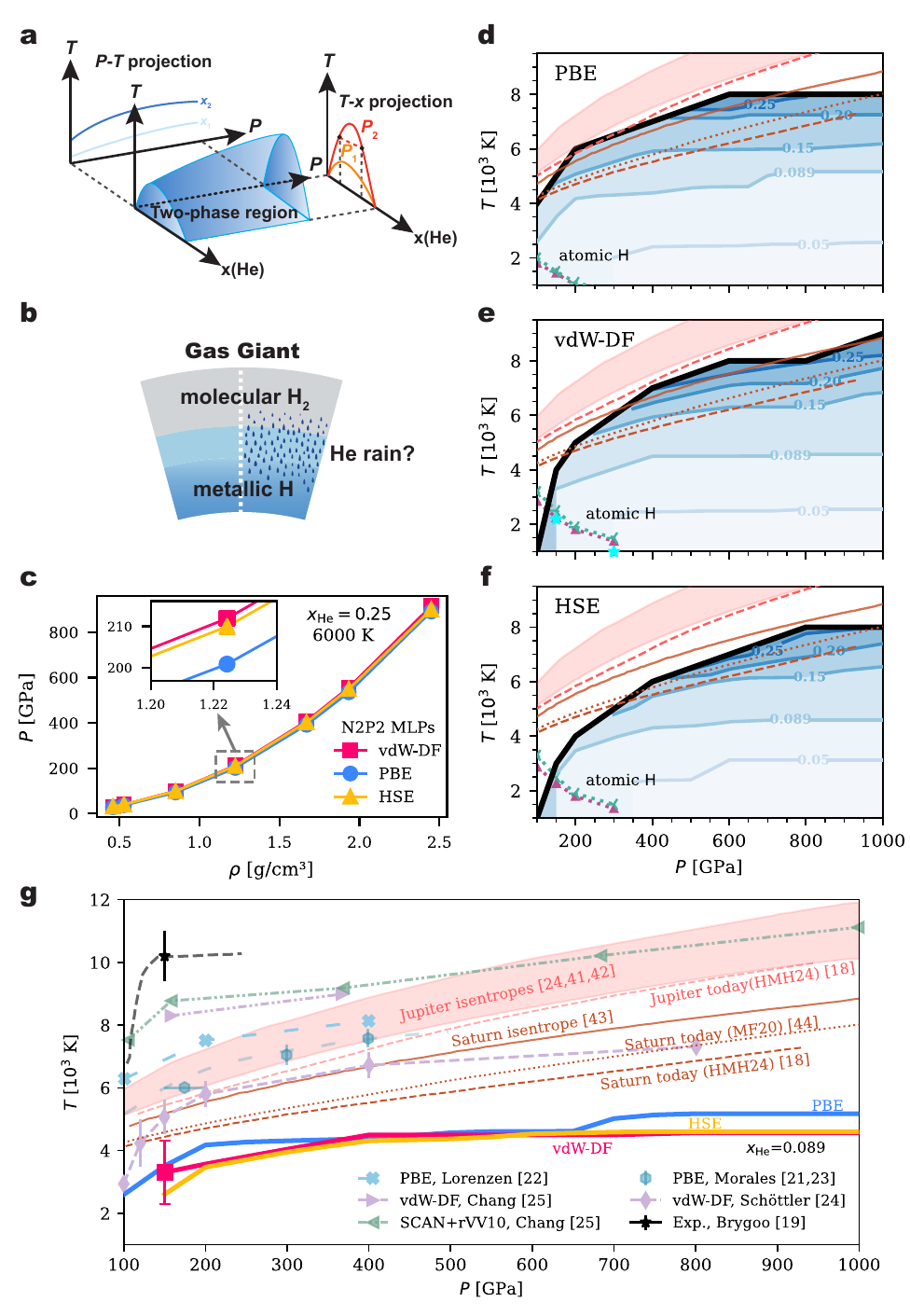}
    \caption{ Immiscibility diagrams of H/He mixtures predicted using MLPs trained on PBE, vdW-DF, and HSE.
    \figLabelCapt{a} Schematic immiscibility diagram of the H/He mixtures with two projections. The blue surface denotes the upper bounds of the two-phase coexistence region. 
    \figLabelCapt{b} A schematic of the possible scenarios of the H/He immiscibility and helium rain in the H/He envelope of gas giants.
    \figLabelCapt{c} Pressure ($P$) as a function of density ($\rho$) for $x_{\rm He}=0.25$ at 6000~K. 
    \figLabelCapt{d, e, f} Composition-dependent immiscibility boundaries of H/He mixtures using MLPs trained on three XC functionals: (d) PBE, (e) vdW-DF, and (f) HSE. The dotted purple and green lines mark the estimated molecular–atomic transition boundaries for pure hydrogen and for a mixture with $x_{\rm He}=0.05$ based on classical MD simulations, respectively. In (e), the cyan stars indicate the transition temperatures obtained from PIMD simulations driven by vdW-DF MLP for $x_{\rm He}=0.05$ at 150 and 300~GPa. Planetary isentropes for Jupiter~\cite{militzer2013ab, nettelmann2008ab, schottler2018ab} (red band indicates the variation among these three studies) and Saturn~\cite{nettelmann2013saturn} (solid brown line) are indicated. 
    The dashed lines denote the present-day thermal profiles of Jupiter and Saturn based on the HMH24 model~\cite{howard2024evolution}, while the dotted brown line denotes another thermal profile of Saturn from the MF20 model~\cite{mankovich2020evidence}. 
    \figLabelCapt{g} H/He immiscibility boundaries at $x_{\rm He}=0.089$. 
    The bar on the red square at 150~GPa is a worst-case estimate of the MLP-induced uncertainty in the vdW-DF phase boundary, approximately $\pm$1000~K, obtained from the FEP analysis of 100 uncorrelated configurations at each $x_{\rm He}$.
    For reference, previous studies at a similar composition ($x_{\rm He} = 0.085\text{–}0.11$) with their reported uncertainties, where available, are included as follows: PBE AIMD results analyzed by assuming ideal entropy of mixing~\cite{lorenzen2011metallization}, PBE AIMD results analyzed with non-ideal entropy~\cite{morales2009, morales2013hydrogen}, vdW-DF AIMD results analyzed with non-ideal entropy~\cite{schottler2018ab}, vdW-DF MLP~\cite{chang2024theoretical}, SCAN+rVV10 MLP~\cite{chang2024theoretical}, and laser-shock experiments~\cite{brygoo2021evidence}.
    }
    \label{Fig:Phase_boundary}
\end{figure*}

The immiscibility boundary is a two-dimensional surface in the $P$–$T$–$x_{\rm He}$ parameter space,
as schematically shown as the blue surface in \figrefsub{Fig:Phase_boundary}{a}.
This boundary can be projected onto the $T$–$x_{\rm He}$ or the $P$–$T$ planes.
A state ($P$, $T$, $x_{\rm He}$) above the surface is fully miscible (e.g., snapshot in \figrefsub{Fig:S0_mu_mixG}{d}). 
Planets with interior conditions all above the immiscibility surface will not undergo H/He phase separation (left side of \figrefsub{Fig:Phase_boundary}{b}).
Conversely, planets with conditions below the surface may support helium rain (right side of \figrefsub{Fig:Phase_boundary}{b}).

We computed $\Delta G^{\rm mix}$ on a dense grid of $P$–$T$ conditions, using the N2P2 MLPs trained on the three XC functionals (PBE, vdW-DF, and HSE): 7 pressures (100, 150, 200, 400, 600, 800, 1000~GPa) and 12 temperatures from 1000~K to 12000~K in 1000~K increments. 
For each set of conditions, H/He phase separation was determined by identifying the point of thermodynamic instability, where $\left(\partial^2 \Delta G^{\rm mix} / \partial x_{\rm{He}}^2\right)_{T, P} = 0$. 
We then constructed continuous immiscibility boundaries at selected $x_{\rm He}$ values by linearly interpolating between the resulting discrete instability points in the $P$–$T$ space. 

We plotted the immiscibility boundaries for various helium fractions, in the $P$–$T$ projection in \figrefsub{Fig:Phase_boundary}{a}. 
The boundaries for PBE, vdW-DF, and HSE are shown in \figrefsub{Fig:Phase_boundary}{d}, \figrefsub{Fig:Phase_boundary}{e}, and \figrefsub{Fig:Phase_boundary}{f}, respectively, with the upper bound indicated by a solid black line. 
Helium isopleths (blue curves) mark the demixing boundary of a given helium fraction, with darker shades corresponding to higher helium fractions. Phase separation occurs below a given isopleth. 

Overall, the computed immiscibility boundaries depend on the XC functional, but the influence is relatively subtle.
As another way to quantify the difference among the three functionals, we compared the EOS in \figrefsub{Fig:Phase_boundary}{c}.  
At $T=6000$~K and $1.225\ \rm{g/cm^3}$, vdW-DF (red square), PBE (blue circle), and HSE (yellow triangle) produce pressures of 212, 201, and 210~GPa, respectively, differing by approximately 5\%. This small variation is consistent with quantum Monte Carlo (QMC) benchmarks~\cite{clay2016benchmarking} that found these three functionals yield pressures that agree within 10\% for H/He mixtures under Jovian conditions.

The immiscibility boundaries from the three functionals have similar characteristics:
The demixing temperature generally increases with pressure and is highly sensitive to the helium fraction, particularly in He-poor compositions. 
This compositional sensitivity is further amplified at higher pressures, evidenced by the greater separation between isopleths in \figrefsub{Fig:Phase_boundary}{d, e, f}. 

We further considered two potential sources that could 
alter the computed boundaries using the MLPs:
First, the difference between the MLP and the underlying DFT might cause a shift.
However, as is seen in \figsrefsub{Fig:S0_mu_mixG}{g, k} and analyzed in detail in the Supplementary Fig.~S15,
such differences lead to very small changes in $\Delta G^{\rm mix}$, so the MLP-derived immiscibility boundaries remain consistent with the DFT reference.
Second, we accounted for the nuclear quantum effects (NQEs) on H/He demixing, by performing path integral MD (PIMD) simulations based on the ring-polymer formalism using the vdW-DF N2P2 MLP.
As shown in the Supplementary Fig.~S28, we found that including NQEs would slightly enlarge the immiscibility region.

The precipitation of helium from the mixture is often regarded as being associated with the metallization of hydrogen~\cite{helled2020understanding}. 
To investigate this connection, we computed the atomic–molecular transition boundaries for both pure hydrogen (dotted purple line) and a H/He mixture with $x_{\rm He} = 0.05$ (dotted green line), as shown in \figsrefsub{Fig:Phase_boundary}{d, e, f}. 
Such a low helium fraction was chosen to quantify the influence of helium on the transition without complications from demixing that make helium concentrations heterogeneous in the system.  
We located the atomic–molecular transition as the $P$–$T$ conditions where half the hydrogen becomes atomic, using an established order parameter~\cite{cheng2020evidence}. 
This order parameter counts the fraction of hydrogen atoms with one bonded hydrogen neighbor, which is computed from H–H RDFs with a smooth cutoff function that is 1 within 0.8~\AA~and decays to 0 at 1.1~\AA. 
Note that the atomic–molecular transition here is a smooth phase transition above the critical point of the first-order transition~\cite{cheng2020evidence}.
Within the H–He immiscibility region, atomic hydrogen dominates, but there is a narrow range of thermodynamic conditions ($T \lessapprox 2000$~K and $P \lessapprox 200$~GPa) where molecular hydrogen is stable.

Across all three functionals, the addition of $x_{\rm{He}}=0.05$ helium to pure hydrogen slightly raises the transition temperature by up to 500~K (\figsrefsub{Fig:Phase_boundary}{d, e, f}), indicating that helium stabilizes molecular hydrogen, in agreement with previous QMC MD~\cite{mazzola2018phase} and PBE AIMD~\cite{Vorberger2007} simulations.
This is because the higher nuclear charge of He tightly localizes electrons and reduces the intermolecular electronic overlap between neighboring hydrogen molecules~\cite{Vorberger2007,mazzola2018phase}.

Moreover, the choice of XC functional affects the predicted H atomic–molecular phase transition.
Both vdW-DF (\figrefsub{Fig:Phase_boundary}{e}) and HSE (\figrefsub{Fig:Phase_boundary}{f}) yield comparable boundaries at $x_{\rm He} = 0.05$ that align with QMC MD results~\cite{mazzola2018phase} at $x_{\rm He} = 0.085$. 
In contrast, PBE (\figrefsub{Fig:Phase_boundary}{d}) gives lower transition temperatures relative to vdW-DF and HSE, likely due to its known underestimation of the hydrogen band gap~\cite{schottler2018ab}.
Detailed comparisons for pure hydrogen and H/He mixtures of the atomic–molecular transition are provided in the Supplementary Figs.~S19 and~S20.

Note that the H atomic–molecular phase transition boundaries discussed above are based on classical MD simulations.
To investigate the impact of NQEs,
we performed PIMD simulations driven by the vdW-DF MLP for $x_{\rm{He}}=0.05$.
As shown by the cyan symbols in \figrefsub{Fig:Phase_boundary}{e},
compared to the classical MD results shown by the green line,
the inclusion of NQEs lowers the hydrogen molecular-to-atomic transition temperature by up to 450~K. 
Such a decrease in the molecular-to-atomic transition temperatures mainly originates from the zero point energy (ZPE) of H–H bond in molecular hydrogen, and is consistent with previous DFT-based~\cite{morales2013nuclear, lu2019towards, hinz2020fully, van2021isotope, bergermann2024nonmetal} and MLP-based PIMD studies~\cite{tenti2025hydrogen}, which report that NQEs reduce the transition temperature by several hundred kelvin.

Our results reveal that the atomic–molecular transition of hydrogen may influence the H/He miscibility diagram: 
According to vdW-DF (\figrefsub{Fig:Phase_boundary}{e}) and HSE (\figrefsub{Fig:Phase_boundary}{f}), hydrogen atomization drastically reduces helium solubility in hydrogen near 150~GPa. 
The mixture with $x_{\rm He} = 0.089$ starts to demix only above this pressure, which is in agreement with previous vdW-DF AIMD results~\cite{schottler2018ab}. 
These trends for vdW-DF and HSE suggest that helium is more soluble in molecular hydrogen than in atomic hydrogen, a mechanism that may promote the H/He demixing in planets~\cite{lorenzen2011metallization, chang2024theoretical}. 
In contrast, because the PBE MLP predicts the atomic–molecular transition at lower temperatures (\figrefsub{Fig:Phase_boundary}{d}), its miscibility diagram is less affected by the transition and exhibits continuous isopleths across the investigated conditions. 
Beyond the atomic–molecular transition ($P>150$~GPa), the three XC functionals predict comparable H/He immiscibility boundaries that level off and increase gradually with pressure. 

To relate our results to planetary models, we show in \figrefsub{Fig:Phase_boundary}{g} the immiscibility boundaries at the protosolar helium abundance ($x_{\rm He} = 0.089$) in the $P$–$T$ projection.
This specific $x_{\rm He}$ serves as the natural baseline because Jupiter and Saturn are often assumed to have formed from the same material as the solar nebula~\cite{bahcall1995solar,helled2020understanding}. 
Previous experimental~\cite{brygoo2021evidence} and theoretical~\cite{lorenzen2011metallization, morales2013hydrogen, schottler2018ab, chang2024theoretical} results for similar compositions ($x_{\rm He} = 0.085\text{–}0.11$) are included for comparison. 

At the protosolar helium abundance, \figrefsub{Fig:Phase_boundary}{g} shows that PBE (solid blue line), vdW-DF (solid red line), and HSE (solid yellow line) predict consistent immiscibility boundaries that differ by less than 500~K.
In detail, at $x_{\rm He} = 0.089$, the PBE MLP predicts the highest demixing temperatures across all pressures, while HSE yields the lowest temperatures at 200~GPa, and vdW-DF gives lower phase-separation thresholds above 600~GPa. 
These trends predicted by the MLPs are in line with previous AIMD studies demonstrating that PBE~\cite{morales2009,morales2013hydrogen} (blue hexagons) produces demixing temperatures up to 1000~K higher than vdW-DF~\cite{schottler2018ab} (purple diamonds).

As illustrated in \figsrefsub{Fig:Phase_boundary}{a, b}, helium rain begins where the internal $P$–$T$ profile of a planet crosses into the two-phase region at the relevant composition.
Accordingly, to investigate the occurrence of helium rain in Jupiter and Saturn, \figsrefsub{Fig:Phase_boundary}{d, e, f} illustrate their $P$–$T$ profiles from two types of planetary models: classical adiabatic models (solid colored lines) and more recent evolutionary models (dotted/dashed colored lines). 
The red band in~\figsrefsub{Fig:Phase_boundary}{d, e, f} reflects an uncertainty of $\sim$1000~K in the isentrope of Jupiter, a range based on three earlier studies~\cite{nettelmann2008ab, militzer2013ab, schottler2018ab}. 
Similarly, to address the uncertainty ($\sim$2000~K) in present-day $P$–$T$ conditions of Saturn in evolutionary models, two references are included: a warmer interior (the MF20 model, dotted brown line)~\cite{mankovich2020evidence} and a cooler interior (the HMH24 model, dashed brown line)~\cite{howard2024evolution}. 
The black curves in~\figsrefsub{Fig:Phase_boundary}{d,e,f} represent the upper boundary of the immiscibility region in the $P$–$T$ projection, while the planetary profiles indicate possible paths through this projected boundary. A crossing of a planetary profile with the boundary marks where the local thermodynamic state enters or exits the two-phase region for the relevant helium concentration.

The formation of helium rain in a gas giant hinges on both its internal thermal profile and helium distribution. 
For Jupiter, the adiabatic models~\cite{militzer2013ab, nettelmann2008ab, schottler2018ab} in \figsrefsub{Fig:Phase_boundary}{d, e, f} place the Jovian isentropes (red bands) largely above the immiscibility boundaries (black curves), suggesting that helium rain is unlikely.
However, the interior of Saturn~\cite{nettelmann2013saturn} (solid brown line) is $\sim$2000~K cooler in the adiabatic model, making helium rain more likely, but whether it occurs would depend on the XC functionals used and the local helium concentration in Saturn: 
PBE (\figrefsub{Fig:Phase_boundary}{d}) predicts helium rain is possible if $x_{\rm He} > 0.20$ under 150–700~GPa, vdW-DF (\figrefsub{Fig:Phase_boundary}{e}) requires a higher helium fraction of $x_{\rm He} > 0.25$ at higher pressures (350–700~GPa), while the HSE MLP predicts no helium rain under the adiabatic conditions.

Recent studies in planetary evolution~\cite{vazan2016evolution,mankovich2020evidence,helled2020understanding,howard2024evolution, arevalo2025jupiter, sur2025simultaneous, bodenheimer2025formation} have challenged the adiabatic assumption for both Jupiter and Saturn and proposed that both planets have cooler interiors shaped by their evolutions. 
These newly developed models incorporate complex processes from the long-term cooling history of planets, including an inhomogeneous helium distribution, formation of helium rain, heavy-element compositional gradients, double-diffusive convection, and a non-adiabatic thermal structure.
We compare our immiscibility boundaries with the thermodynamic conditions predicted by these models (dotted/dashed lines in \figsrefsub{Fig:Phase_boundary}{d–g}).
For Jupiter, the interior profile predicted by the HMH24 model~\cite{howard2024evolution} (dashed red lines) lies above the immiscibility boundaries at the protosolar helium abundance. 
Although helium rain may occur for $x_{\rm He} > 0.15$ (\figrefsub{Fig:Phase_boundary}{d}), recent planetary models~\cite{mankovich2020evidence, howard2024evolution, arevalo2025jupiter, sur2025simultaneous} suggest this composition requirement is not met in Jupiter, where the peak helium abundance is estimated to be $x_{\rm He} = 0.09$, indicating that helium rain is unlikely in Jupiter.

On the contrary, the $P$–$T$ conditions of Saturn (dashed brown line) from the same HMH24 model~\cite{howard2024evolution} and the warmer MF20 profile (dotted brown line) from an alternative evolutionary model~\cite{mankovich2020evidence} both fall below the critical boundaries (black lines in \figsrefsub{Fig:Phase_boundary}{d, e, f}). For all functionals, helium rain requires $x_{\rm He} \geq 0.15$, as shown in \figsrefsub{Fig:Phase_boundary}{d, e, f}.
Given that Saturn may have a large helium gradient with $x_{\rm He}$ varying from 0.1 up to 1.0 based on recent studies~\cite{mankovich2020evidence, howard2024evolution, sur2025simultaneous,markham2024stable, bodenheimer2025formation, sur2026next}, helium rain may occur throughout the interior of Saturn.
However, this inference remains sensitive to the adopted models and the resulting thermal profile of Saturn.

Our H/He immiscibility boundaries (solid lines in \figrefsub{Fig:Phase_boundary}{g}) are systematically shifted to lower temperatures by approximately 2000~K for $x_{\rm He} = 0.089$, compared to prior AIMD studies using PBE~\cite{lorenzen2011metallization, morales2013hydrogen} (blue-coded symbols) and vdW-DF~\cite{schottler2018ab} (purple diamonds).  
According to our PBE MLP (solid blue line), at the protosolar helium abundance, demixing occurs only at much lower temperatures, such as 3511~K at 150~GPa. 
In contrast, prior PBE AIMD analyses reported demixing at 7500~K and 160~GPa by assuming ideal entropy of mixing~\cite{lorenzen2011metallization} (blue cross), and at 6000~K and 175~GPa including non-ideal effects on entropy~\cite{morales2013hydrogen} (blue hexagon). 
We will explore the possible origins of this discrepancy in detail in the next section.
A recent large-scale MLP MD study~\cite{chang2024theoretical} predicted even higher demixing temperatures, which are shown by the purple (vdW-DF) and green (SCAN+rVV10) triangles in \figrefsub{Fig:Phase_boundary}{g}. 
While their immiscibility boundaries are robust with respect to the chosen functionals, the demixing temperatures are elevated by $\sim$2000~K relative to previous AIMD studies and by $\sim$4000~K relative to our predictions. 
A fundamental difference in methodology may contribute to this discrepancy: Whereas our method and prior AIMD studies compute the composition-dependent $\Delta G^{\rm mix}$, their approach defines phase separation in terms of changes in local atomic configurations.

\subsection{Thermodynamic models of H/He mixtures}

To rationalize the driving force behind H/He phase separation and understand the discrepancies with previous results, we fitted $\Delta G^{\rm mix}$ to a Redlich–Kister (R–K) solution model~\cite{redlich1948thermodynamics}.
While previous AIMD studies~\cite{lorenzen2011metallization, schottler2018ab} employed fifth-order R–K models for $\Delta G^{\rm mix}$, here we adopted a third-order form, because including higher-order terms $\omega_i(T, P)$ with $i$ = 4, 5 was found to provide no improvement in either fit accuracy or predictive power (see the Supplementary Figs.~S23 and~S24 for the 5th-order fits and related analyses). 
Specifically, the solution model used is expressed as: 
\begin{equation}
\begin{aligned}
    \Delta G^{\rm mix}  & =  k_B T ( { x_{\rm{He}} \ln  x_{\rm{He}} + x_{\rm{H}} \ln x_{\rm{H}}  })  + \\
                        & x_{\rm{He}}x_{\rm{H}} \sum_{i=1}^{3} \omega_{i} \left(1-2 x_{\rm{He}}\right)^{i-1} , 
\end{aligned}
\label{Eq: R–K model}
\end{equation}
where the parameters $\omega_i$ capture non-ideal interactions that lead to phase separation.

In accordance with the assumption of a homogeneous mixture in the R–K model, we fitted $\omega_i(T, P)$ parameters only to single-phase data above the immiscibility boundary, as shown in \figrefsub{Fig:Phase_boundary}{a}. Results based on vdW-DF are exemplified in \figrefsub{Fig:Thermodynamic_model}{a}, with PBE and HSE fits detailed in the Supplementary Figs.~S21 and~S22. 
The suitability of the R–K model for high-pressure H/He mixtures is demonstrated by the close alignment between the R–K fits (solid lines) and the direct MD results (colored pentagons). 
For the two-phase region, where the real free energy curve reduces to a straight line (e.g., as illustrated in \figrefsub{Fig:S0_mu_mixG}{k}), we used dashed lines to denote the extrapolated $\Delta G^{\rm mix}$ values for homogeneous mixtures from the individual R–K fits. 
At lower temperatures and higher pressures, however, the wide miscibility gap with linear mixing free energy (\figrefsub{Fig:S0_mu_mixG}{k}) leaves insufficient data for an accurate local fit. 
To overcome this limitation, we performed a global fit of $\Delta G^{\rm mix}$ to parameterize the $P$–$T$ dependence of $\omega_i$ and extract their coefficients to model $\Delta G^{\rm mix}$, which is shown as dotted lines in \figrefsub{Fig:Thermodynamic_model}{a}. 
Specifically, $\omega_1(T, P)$ and $\omega_2(T, P)$ parameters were fitted to the polynomial expression $\omega_{i} = c + a_1 T + a_2 T^2 + b_1 P + b_2 P^2 $, with fits and uncertainties for vdW-DF presented in \figrefsub{Fig:Thermodynamic_model}{b}.
We treated $\omega_3(T, P)$ as constant for a given pressure, since it displays no discernible trends with temperature. 

As shown in \figrefsub{Fig:Thermodynamic_model}{b}, the $\Delta G^{\rm mix}$ values and hence the thermodynamic stability of H/He mixtures are primarily controlled by the $\omega_1$ term. 
Across all investigated conditions, $\omega_1$ and $\omega_3$ remain positive, increasing $\Delta G^{\rm mix}$ and destabilizing the H/He mixture. 
With increasing temperature, the decrease of $\omega_1$ lowers $\Delta G^{\rm mix}$ of a homogeneous mixture and enhances H/He miscibility. 
The contribution of $\omega_2$ to $\Delta G^{\rm mix}$ depends on both composition and temperature; see the second panel of~\figrefsub{Fig:Thermodynamic_model}{b}. 
Below 6000~K, the second-order term $\omega_2 x_{\rm{He}}x_{\rm{H}}(1-2x_{\rm{He}})$ in Eq.~\ref{Eq: R–K model} increases $\Delta G^{\rm mix}$ for He-poor compositions ($x_{\rm He} < 0.5$), but decreases $\Delta G^{\rm mix}$ for He-rich compositions ($x_{\rm He} > 0.5$). 
Above 6000~K, $\omega_2$ becomes negative and the trend reverses: the second-order term now lowers $\Delta G^{\rm mix}$ for He-poor components and increases it for He-rich components. 
Therefore, these trends indicate that H/He demixing is mainly driven by $\omega_1$ and $\omega_3$ at low temperatures, while higher temperatures and the sign reversal of $\omega_2$ progressively favor complete miscibility, particularly for He-poor compositions.

Next, we used the fitted R–K model to analytically calculate the H/He immiscibility boundary for $x_{\rm He}=0.089$ by solving the $P$–$T$ conditions that satisfy the equation $\left(\partial^2 \Delta G^{\rm mix} / \partial x_{\rm{He}}^2\right) = 0$. 
The resulting boundaries (dashed lines in \figrefsub{Fig:Thermodynamic_model}{c}) agree well with our direct simulation results (solid lines).
The uncertainty in the immiscibility boundary (shaded regions) is quantified via Monte Carlo sampling that propagates the error in the fitted parameters $\omega_i$ to $\Delta G^{\rm mix}$ according to Eq.~\ref{Eq: R–K model}.
The choice of functional has little impact on the results, as the predictions from different XC functionals fall within the uncertainty range of thermodynamic models. 
This analysis also shows that fitting the R–K model alone introduces an uncertainty of approximately 1000~K in the predicted phase boundary compared to direct MD results, due to uncertainties in the polynomial coefficients.

We then investigated the possible explanation for the $\sim$2000~K discrepancy between our H/He immiscibility boundaries and previous AIMD results~\cite{lorenzen2011metallization,morales2009, morales2013hydrogen, schottler2018ab}. 
First, we examined the effect of the thermodynamic model fitting protocol: 
In contrast to our global-fit approach described above, previous AIMD work~\cite{morales2013hydrogen, schottler2018ab} used an individual fitting scheme: 
They performed individual R–K fits to $\Delta G^{\rm mix}$ at each $P$–$T$ condition, and used common-tangent analyses of the modeled $\Delta G^{\rm mix}$ to identify phase separation and the compositions of any demixed phases. 
At a fixed pressure, these demixing temperatures were combined with the corresponding $x_{\rm He}$ of the demixed phases to map the $T$–$x_{\rm He}$ immiscibility diagram (right side of \figrefsub{Fig:Phase_boundary}{a}).
By treating the demixing temperature as a function of $x_{\rm He}$, the demixing temperature at the protosolar helium abundance was linearly interpolated in the $T$–$x_{\rm He}$ projection. 
Applying the same protocol to our vdW-DF MLP MD results at 150~GPa and 200~GPa produces immiscibility boundaries (pink stars in \figrefsub{Fig:Thermodynamic_model}{c}) that agree with our global-fit results (dashed red lines) within 500~K. 
The same agreement holds for the PBE and HSE results (detailed in the Supplementary Table~S8), confirming the validity of both the global-fit and individual-fit approaches.
Moreover, the inferred boundary is insensitive to the order of R–K models used; the fifth-order R–K fit yields immiscibility boundaries consistent with the third-order fit. 
Therefore, methodological differences in fitting and post-analysis procedures between our work and prior AIMD studies~\cite{lorenzen2011metallization, schottler2018ab} can only account for a shift of at most $\sim$1000~K (see \figrefsub{Fig:Thermodynamic_model}{c} and the Supplementary Fig.~S24). 

As such, the $\sim$2000~K discrepancy cannot be attributed to R–K model fitting but can only originate from the computed $\Delta G^{\rm mix}$ itself.
Indeed, direct comparisons of $\Delta G^{\rm mix}$ from MLP MD and AIMD based on vdW-DF at 400~GPa and 1000~GPa over 3000–10000~K show substantial deviations, irrespective of whether non-ideal effects in entropy are included in the analysis of AIMD results, as detailed in the Supplementary Fig.~S25.
We further speculate that the discrepancy in computed $\Delta G^{\rm mix}$ may stem from the system size in the modeling of the mixtures.
Our large-scale MD simulations allow macroscopic phase separation (\figrefsub{Fig:S0_mu_mixG}{h}), and produce a demixed, linear segment in the derived $\Delta G^{\rm mix}$ (\figrefsub{Fig:S0_mu_mixG}{k}).
In contrast, earlier AIMD studies~\cite{morales2009, lorenzen2011metallization, morales2013hydrogen, schottler2018ab} use a relatively small system ($N\leq 64$~atoms) to enforce complete mixing at all conditions. 
The single-phase assumption enables the evaluation of $\Delta G^{\rm mix}$ for a homogeneous H/He mixture, analogous to our hypothetical green curve in \figrefsub{Fig:S0_mu_mixG}{k}.
As $\Delta G^{\rm mix}$ comprises enthalpic and entropic contributions, the enthalpy is directly accessible from AIMD calculations, and the entropy is either taken as the ideal entropy of mixing (e.g.,~\cite{lorenzen2011metallization}), or evaluated by integrating changes in thermodynamic properties (e.g., internal energy and partial molar volume) over EOSs starting from a reference entropy for each $x_{\rm{He}}$ (e.g.,~\cite{morales2009, morales2013hydrogen, schottler2018ab}). 
This reference entropy can be computed via coupling-constant integration with force matching~\cite{morales2009, schottler2018ab}. 
However, if the small system is not perfectly mixed, or if the free energy of the finite-size system deviates from larger systems at the thermodynamic limit, errors can propagate into the calculated $\Delta G^{\rm mix}$ and the inferred immiscibility boundary. In the Supplementary Figs.~S9–S12, we examined finite-size effects and showed that the complete-mixing assumption may be violated even in small systems.

\section{Discussion}
By analyzing the thermodynamic behaviors of H/He mixtures with $\Delta G^{\rm mix}$ computed from the S0 method, as shown in \figref{Fig:S0_mu_mixG}, we reported immiscibility diagrams of hydrogen and helium under planetary conditions in \figref{Fig:Phase_boundary} using three MLPs trained on different XC functionals: PBE, vdW-DF, and HSE. 
The derived H/He immiscibility boundary (\figsrefsub{Fig:Phase_boundary}{d, e, f}) strongly depends on temperature and helium atomic fraction while exhibiting comparatively less sensitivity to the XC functional, consistent with previous AIMD and MLP MD studies~\cite{lorenzen2011metallization, morales2009, morales2013hydrogen, schottler2018ab,chang2024theoretical}. 
The immiscibility boundaries reported here should be interpreted within the accuracy of the underlying Born–Oppenheimer DFT framework, MLP fitting, MLP-driven classical MD, and the numerical free-energy calculations. One systematic uncertainty arises from the exchange-correlation functional (PBE, vdW-DF, and HSE predict immiscibility boundaries that differ by less than 500~K at the protosolar helium abundance). Meanwhile, the MLP-to-DFT free-energy perturbation tests indicate the MLP error is not important: MLP and underlying DFT predict statistically indistinguishable mixing free energies at temperatures relevant to the immiscibility boundary, and MLP errors would lead to a shift of less than 1000~K in the worst-case scenario analysis (see the Supplementary Information Sections IV and VII). The PIMD simulations indicate that NQEs would only slightly affect the immiscibility boundaries by less than 200~K. Thus, although the absolute demixing temperatures remain condition-dependent and uncertain at the level of several hundred kelvin, the lowering of the H/He immiscibility boundary relative to previous small-cell AIMD estimates is robust across the three exchange-correlation functionals tested here.
Moreover, the close match between immiscibility boundaries derived from the fitted R–K models (\figref{Fig:Thermodynamic_model}) and from direct MD simulations demonstrates the reliability of the thermodynamic model. This can make it a robust tool for the modeling of mass transport driven by chemical potential gradients in planets. 
Overall, as shown in \figrefsub{Fig:Phase_boundary}{g}, our immiscibility boundaries shift to lower temperature regions compared to laser-shock experimental measurements~\cite{brygoo2021evidence} and previous theoretical results~\cite{lorenzen2011metallization, morales2013hydrogen, schottler2018ab, chang2024theoretical}.

Relating our results to planetary models reveals distinct scenarios for helium rain in Jupiter and Saturn, as sketched in \figrefsub{Fig:Phase_boundary}{b}. 
As for Jupiter, both adiabatic and evolutionary models indicate helium rain is unlikely: 
First, the isentrope of Jupiter lies above the immiscibility boundary (\figsrefsub{Fig:Phase_boundary}{d, e, f}), implying that hydrogen and helium remain fully mixed. 
Second, although the latest evolutionary models (e.g., HMH24~\cite{howard2024evolution}) predict the interior of Jupiter to be about 1000~K cooler (dashed red lines in \figsrefsub{Fig:Phase_boundary}{d, e, f}), the predicted peak helium abundance of $x_{\rm He} \approx 0.09$~\cite{mankovich2020evidence, howard2024evolution, arevalo2025jupiter, sur2025simultaneous} is still below the concentration threshold for helium rainout ($x_{\rm He}\approx0.15$) under the relevant conditions. 
Therefore, Jupiter has likely not cooled sufficiently for helium rain to initiate, and alternative mechanisms are required to explain its atmospheric helium depletion. 

On the contrary, the cooler interior of Saturn predicted by the evolutionary models~\cite{howard2024evolution} (dotted/dashed brown lines in~\figsrefsub{Fig:Phase_boundary}{d–f}) may support extensive helium rain that potentially reaches the deeper interior ($P\approx1000$~GPa).  
The substantial overlap between the present $P$–$T$ condition of Saturn from HMH24/MF20 and the two-phase region, as illustrated in \figsrefsub{Fig:Phase_boundary}{d–f}, suggests that helium rain may have begun in Saturn long ago and could persist over an extended period of time.
This is because demixing starts when the planetary thermal profile first intersects the immiscibility boundary, drives helium redistribution and modifies the thermodynamic conditions in the planet, and ceases once the evolved planetary profile no longer lies within the two-phase domain. 
This ongoing helium rain can help explain several unique characteristics of Saturn: 
The release of latent heat and gravitational energy during helium precipitation serves as an additional heat source in Saturn, which could account for its anomalous luminosity~\cite{stevenson1977dynamics, fortney2011self}.
Gravitational differentiation transports denser He-rich droplets inward and would explain the observed atmospheric helium depletion in Saturn~\cite{young2003galileo}.
The resulting stratification could establish a compositional gradient in Saturn that shapes its dynamo processes and axisymmetric magnetic field~\cite{mankovich2020evidence, howard2024evolution, pustow2016h, preising2023material, helled2020understanding}. 
As our calculated immiscibility boundaries shift to lower temperatures than those in previous studies~\cite{lorenzen2011metallization, morales2013hydrogen, schottler2018ab, brygoo2021evidence, chang2024theoretical}, we predict a delayed onset of H/He demixing during the cooling history of Saturn. 

The composition-dependent immiscibility boundaries derived here establish constraints for interpreting the distinct evolution and structures of Jupiter and Saturn~\cite{stevenson1977dynamics, fortney2011self}.
Once H/He demixing is triggered, a gas giant can no longer be modeled as a homogeneous adiabatic planet~\cite{helled2020understanding, mankovich2020evidence, mankovich2021diffuse, howard2024evolution}.
Detailed $P$–$T$–$x_{\rm He}$ conditions for H/He immiscibility are thus critical inputs to future planetary studies to model the magnitude of helium rain, atmospheric helium abundances, planetary heat budgets, emergent compositional gradients, and associated fluid dynamics. 
Beyond the solar system, resolving the phase behaviors of H/He mixtures and their role in the evolution pathways is fundamental to characterizing exoplanetary gas giants~\cite{muller2023warm}.

\section{Methods}

\subsection{DFT calculations}
\label{Method:DFT}
We used the plane-wave code VASP~\cite{kresse1996efficient} for production DFT calculations using the exchange-correlation functionals of PBE~\cite{perdew1996generalized}, vdW-DF~\cite{lee2010higher}, and HSE~\cite{heyd2003hybrid, heyd2006hybrid}. 
We sampled the Brillouin zone using a $k$-point grid resolution of at least 0.02~\AA$^{-1}$ for the PBE and vdW-DF calculations. 
The HSE calculations were performed using a resolution of 0.03~\AA$^{-1}$. We note that for the $\Delta$-learning from the vdW-DF baseline to HSE, the vdW-DF energies and forces are recalculated at a resolution matching the HSE calculations (0.03~\AA$^{-1}$).

The plane-wave energy cutoff was set to 1200~eV, after performing convergence tests for energy, forces, and pressure across a 700–1300~eV range. For the pseudopotentials, we used the hard version of the hydrogen PBE projector augmented wave (PAW) potential (\verb|PAW_PBE H_h 06Feb2004|), while for helium we used the only available version (\verb|PAW_PBE He 05Jan2001|) from the VASP library. 

In the end, the DFT-generated training set for the PBE and vdW-DF functionals comprised 13,389 configurations (1,553,967 atoms) with varying $x_{\rm He}$ across a broad range of $P$–$T$ conditions (1–1200~GPa and 500–16000~K). 
Subsets of approximately 2,000 configurations each were used in fitting $\Delta$-learning MLPs for vdW-DF and HSE. 

In addition, we generated a dataset of 3,500 configurations using vdW-DF DFT with the system size of $N=64$ for three $P$–$T$ conditions (150~GPa, 2000~K; 150~GPa, 5000~K; 800~GPa, 7000~K) and 12 helium compositions,
as well as another set of 3,500 configurations using vdW-DF DFT with the system size of $N=1024$ at the same $P$–$T$–$x_{\rm He}$ conditions.
These two datasets were used to assess finite-size effects in the DFT reference data and to further validate the fitted MLPs.

\subsection{Construction of machine learning potentials}
To construct MLPs for the high-pressure H/He system, we employed the Behler–Parrinello neural network~\cite{behler2007generalized} via the N2P2 code~\cite{singraber2019parallel} on CPUs, and the equivariant message-passing MACE~\cite{batatia2022mace} and CACE~\cite{cheng2024cartesian} models on GPUs. 

In total, we developed 8 MLPs using three MLP architectures and two training strategies for three XC functionals (PBE, vdW-DF, HSE). 
Six MLPs were obtained by direct fitting: one for each of the three architectures for both PBE and vdW-DF. 
The other two MLPs were developed using a $\Delta$-learning strategy in the N2P2 package: a vdW-DF MLP was trained on the $\Delta$ component $\Delta$(PBE$\rightarrow$vdW-DF) relative to the PBE MLP baseline, and an HSE MLP was trained on $\Delta$(vdW-DF$\rightarrow$HSE) relative to the vdW-DF MLP baseline. 
We evaluated the performance of all MLPs for H/He mixtures by comparing energetic and thermodynamic properties against AIMD results and one another.
Dataset construction and training procedures are detailed in Supplementary Information Section~II, with model errors reported in Supplementary Tables~S1–S2; the structural and EOS benchmarks are shown in Supplementary Figs.~S3–S8.

Moreover, we compared the performance of these models with MLPs trained using alternative fitting protocols, including models trained with stress data and models trained on restricted subsets of the dataset. 
These comparisons further validate the training strategies and model architectures of our MLPs, as reported in Supplementary Tables~S2 and~S3.

\subsection{MLP MD simulation details}
All MD simulations for N2P2 MLPs were performed in LAMMPS~\cite{LAMMPS} on CPUs. 
Following a 10~ps pre-equilibration for a 3456-atom system, each production run was performed for 60~ps with a timestep of 0.2~fs in the isothermal–isobaric (NPT) ensemble using the Nos\'e–Hoover thermostat and barostat. 
For MLPs constructed with MACE and CACE, MD simulations were conducted with the ASE package~\cite{larsen2017atomic} utilizing the Nos\'e–Hoover thermostat and Berendsen barostat. 
Sensitivity analyses on the timestep (0.1–0.2~fs) and simulation duration (40–80~ps) showed no discernible quantitative differences in computed chemical potentials and free energies. Further details are provided in the Supplementary Fig.~S13.

\subsection{Chemical potentials of H/He mixtures}
We applied the newly established S0 method~\cite{cheng2022computing} to compute the chemical potentials of H/He mixtures. 
For such a purpose, structure factors $S(\mathbf{k})$ were sampled every 10~fs from equilibrium MD trajectories at each $P$–$T$–$x_{\rm He}$ state. In the limit of wave-vector $\mathbf{k} \rightarrow 0$, the static structure factor for the H–He pair, $S_{\ce{H-He}}^0$, is derived as:
\begin{equation}
\begin{split}
S_{\ce{H-He}}^0 &= \lim_{\mathbf{k} \rightarrow 0} S_{\ce{H-He}}(\mathbf{k})  \\ 
&= \lim_{\mathbf{k} \rightarrow 0} \frac{\langle\sum_{i}^{N_{\ce{H}}} \exp \left(i \mathbf{k} \cdot \mathbf{r}_{i}(t)\right) \sum_{j}^{N_{\ce{He}}} \exp \left(-i \mathbf{k} \cdot \mathbf{r}_{j}(t)\right) \rangle}{\sqrt{N_{\ce{H}} N_{\ce{He}}}}.
\end{split}
\end{equation}
In particular, $S_{\ce{H-He}}^0$ was obtained by fitting small-$\mathbf k$ $S_{\ce{H-He}}(\mathbf{k})$ to the Ornstein–Zernike form~\cite{ben2006molecular} ($S_{\ce{H-He}}(\mathbf k) = {S_{\ce{H-He}}^0} / (1+ a k^2)$) with a cutoff of $k^2_{\rm{cut}}= 0.01 \times 4\pi^2/\mathrm{\mathring{A}^2}$.
By running multiple MD simulations at varying $x_{\rm He}$, the chemical potential of helium ($\mu_{\rm He}$) was calculated through numerical integration based on Eq.~\ref{Eq: mu_i_liq_state}:
\begin{equation}
\begin{split}
\mu_{\rm He}(T, P, x_{\rm He})  & = \mu_{{\rm He}, \rm{pure}} (T, P)   \\
    & \quad +  k_{\mathrm{B}} T \int_{0}^{\ln x_{\rm He}} d (\ln x_{\rm He} ) \left(\frac{\partial \mu_{\rm He}}{\partial \ln x_{\rm He}}\right)_{T, P},  
\end{split}
\end{equation}
where the chemical potential of a pure helium liquid ($\mu_{{\rm He},\rm{pure}} (T, P)$) was selected as a baseline for convenience, which does not affect the immiscibility diagram. 
Accordingly, the chemical potentials for hydrogen ($\mu_{\rm H}$) were derived based on the Gibbs–Duhem equation~\cite{cheng2022computing}. 

Equilibrium MD simulations were performed across a grid of $P$–$T$–$x_{\rm He}$ conditions: $P$ set at 100, 150, 200, 400, 600, 800, and 1000~GPa, $T$ ranging from 1000 to 12000~K in increments of 1000~K, and $x_{\rm He}$ varying from 0.015 to 0.98 in 26 intervals.
Using the S0 method, chemical potentials at each $P$–$T$ condition for each XC functional were computed and are shown in Supplementary Figs.~S16–S18, and the corresponding miscibility gaps are listed in Supplementary Tables~S4–S6.

\subsection{Promoting MLP-derived $\Delta G^{\rm mix}$ to the DFT level}

To quantify the effect of the MLP error, we corrected the MLP-derived $\Delta G^{\rm mix}$ to the DFT level using free-energy perturbation (FEP) at three $P$–$T$ conditions (150~GPa, 2000~K; 150~GPa, 5000~K; 800~GPa, 7000~K). 
For each $P$–$T$–$x_{\rm He}$ condition, we sampled 100 uncorrelated configurations of system size $N=1024$ from MD simulations using the vdW-DF N2P2 MLP. 
We recomputed the vdW-DF DFT energies and forces for these configurations, 
which constitute the $N=1024$ vdW-DF DFT dataset described above.
We then computed the energy difference, $\Delta U = U_{\rm DFT} - U_{\rm MLP}$, where $U_{\rm DFT}$ is the DFT energy of a given configuration, and $U_{\rm MLP}$ is the corresponding MLP energy evaluated for the same configuration. The free-energy difference per atom for the system at the DFT and the MLP level, $\Delta g$, can then be estimated as:
\begin{equation}
N \Delta g = -k_{\rm B} T \ln\left\langle \exp\left(-\frac{\Delta U}{k_{\rm B} T}\right) \right\rangle_{\rm MLP}.
\end{equation}
Using such free energy correction terms $\Delta g$, we promoted the MLP-derived $\Delta G^{\rm mix}$ to the DFT level for systems at fixed $P$–$T$–$x_{\rm He}$.

\subsection{PIMD simulation details}
To account for NQEs in H/He mixtures, we performed PIMD simulations~\cite{tuckerman2023statistical} based on the ring-polymer formalism via the i-PI interface (version 3.0)~\cite{ceriotti2014pi, litman2024pi} coupled with LAMMPS, using the vdW-DF N2P2 MLP. 
Following prior convergence benchmarks~\cite{hinz2020fully, van2021isotope}, we employed eight beads over the investigated temperature range of 1000–3000~K. 
Consistent with earlier PIMD studies of dense hydrogen~\cite{van2021isotope}, each PIMD simulation began with a 10~ps equilibration in the NPT ensemble, followed by an 80~ps production run in the NVT ensemble. 
During the NPT equilibration stage, pressure was controlled with the isotropic stochastic barostat using a relaxation time of 500~fs. 
Temperature was controlled using the stochastic path integral Langevin equation thermostat (\texttt{PILE\_L}) in i-PI~\cite{ceriotti2010efficient}, with a relaxation time of 50~fs. 
A system size of $N=2560$ and a timestep of 0.2~fs were selected to ensure numerical convergence while maintaining computational efficiency.
Further details are provided in the Supplementary Information Section VII.

\textbf{Data availability}
The training data, fitted MLPs, MLP molecular dynamics results, and source data underlying all figures and tables in the main text and Supplementary Information have been deposited in the Zenodo repository: \url{https://doi.org/10.5281/zenodo.21632034}~\cite{wang_2026_21632034}.

\textbf{Code availability}
The data-analysis and plotting scripts used to generate the figures and tables in the main text and Supplementary Information are available in the GitHub repository: \url{https://github.com/ChengUCB/HighPressure_HHe}.

\textbf{Acknowledgments}
The authors are thankful to Armin Bergermann and Ronald Redmer for generously sharing AIMD trajectories and data, and for valuable discussions and feedback on the manuscript. 
The authors also acknowledge the research computing facilities provided by the high-performance computing (HPC) cluster at the Institute of Science and Technology Austria (ISTA), Berkeley Research Computing (BRC) at the University of California, Berkeley (UCB), and Lawrence Livermore National Laboratory (LLNL).

\textbf{Funding Statement}
B.C. was supported by the Laboratory Directed Research and Development (LDRD) Program at Lawrence Berkeley National Laboratory (LBNL) under project ID 110990-001.
Part of this work was performed under the auspices of the U.S. Department of Energy by LLNL under Contract DE-AC52-07NA27344. S.H. was supported by the LDRD Program at LLNL under project tracking code 23-SI-006.

\textbf{Author Contributions Statement}
B.C. conceived the idea; B.C. and X.W. designed the research; X.W., S.H., and B.C. performed the research; X.W., S.H., and B.C. wrote the paper.

\textbf{Competing Interests Statement}
B.C. has an equity stake in AIMATX Inc. The remaining authors declare no other competing interests.

\textbf{Figure Legends/Captions}

\begin{figure*}[htbp!]
    \centering
   \includegraphics[width=\textwidth]{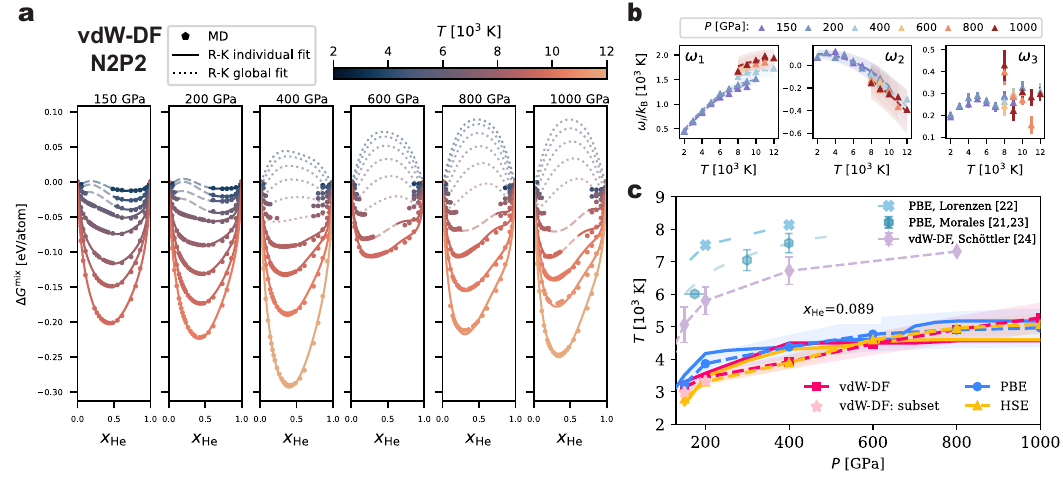}
    \caption{
    Redlich–Kister (R–K) model analyses of the mixing free energy ($\Delta G^{\rm mix}$) and the resulting immiscibility boundary in the H/He system.
    \figLabelCapt{a} Mixing free energy per atom of H/He mixtures computed from 150~GPa to 1000~GPa and 2000~K to 12000~K, using the vdW-DF MLP. 
    The solid lines denote the third-order R–K model fitted individually at each $P$–$T$. The dashed lines indicate the modeled $\Delta G^{\rm mix}$ for assumed homogeneous H/He mixtures in the phase separation regime, derived from the individual fits. The dotted lines show $\Delta G^{\rm mix}$ derived from a global R–K fit. 
    \figLabelCapt{b} Pressure and temperature dependence of the R–K parameters ($\omega_i$). Triangle symbols show $\omega_i$ values obtained from individual fits at each $P$–$T$ state point, with error bars indicating $\pm1$ standard error. For $\omega_1$ and $\omega_2$, the dashed lines represent the global polynomial fits in $P$ and $T$ to MD-derived values at 32 $P$–$T$ state points with shaded bands indicating $\pm1$ standard error propagated from the fitted-coefficient variances. 
    \figLabelCapt{c} Comparison of H/He immiscibility boundaries from different methods at a helium fraction of $x_{\rm He}=0.089$. 
    The blue, red, and yellow solid lines correspond to direct MD simulation results based on PBE, vdW-DF, and HSE MLPs, respectively. 
    The corresponding dashed lines show the immiscibility boundaries predicted by the globally fitted R–K model, and the shaded regions indicate the $\pm1$ standard deviation across 500 Monte Carlo samples generated by propagating the fitting uncertainties of $\omega_i$ values into the derived phase boundaries. 
    The pink stars represent the immiscibility boundaries derived with the methodology of individual $P$–$T$ fit in earlier AIMD studies~\cite{morales2013hydrogen, schottler2018ab} for a vdW-DF subset at 150~GPa and 200~GPa.
    Previous studies at similar compositions ($x_{\rm He} = 0.085\text{–}0.09$) are included: PBE AIMD results analyzed by assuming ideal entropy of mixing~\cite{lorenzen2011metallization} (blue crosses), PBE AIMD results analyzed with non-ideal entropy~\cite{morales2009, morales2013hydrogen} (blue hexagons), vdW-DF AIMD results with non-ideal entropy~\cite{schottler2018ab} (purple diamonds).
    }
    \label{Fig:Thermodynamic_model}
\end{figure*}


\begin{thebibliography}{79}%
\makeatletter
\providecommand \@ifxundefined [1]{%
 \@ifx{#1\undefined}
}%
\providecommand \@ifnum [1]{%
 \ifnum #1\expandafter \@firstoftwo
 \else \expandafter \@secondoftwo
 \fi
}%
\providecommand \@ifx [1]{%
 \ifx #1\expandafter \@firstoftwo
 \else \expandafter \@secondoftwo
 \fi
}%
\providecommand \natexlab [1]{#1}%
\providecommand \enquote  [1]{``#1''}%
\providecommand \bibnamefont  [1]{#1}%
\providecommand \bibfnamefont [1]{#1}%
\providecommand \citenamefont [1]{#1}%
\providecommand \href@noop [0]{\@secondoftwo}%
\providecommand \href [0]{\begingroup \@sanitize@url \@href}%
\providecommand \@href[1]{\@@startlink{#1}\@@href}%
\providecommand \@@href[1]{\endgroup#1\@@endlink}%
\providecommand \@sanitize@url [0]{\catcode `\\12\catcode `\$12\catcode `\&12\catcode `\#12\catcode `\^12\catcode `\_12\catcode `\%12\relax}%
\providecommand \@@startlink[1]{}%
\providecommand \@@endlink[0]{}%
\providecommand \url  [0]{\begingroup\@sanitize@url \@url }%
\providecommand \@url [1]{\endgroup\@href {#1}{\urlprefix }}%
\providecommand \urlprefix  [0]{URL }%
\providecommand \Eprint [0]{\href }%
\providecommand \doibase [0]{http://dx.doi.org/}%
\providecommand \selectlanguage [0]{\@gobble}%
\providecommand \bibinfo  [0]{\@secondoftwo}%
\providecommand \bibfield  [0]{\@secondoftwo}%
\providecommand \translation [1]{[#1]}%
\providecommand \BibitemOpen [0]{}%
\providecommand \bibitemStop [0]{}%
\providecommand \bibitemNoStop [0]{.\EOS\space}%
\providecommand \EOS [0]{\spacefactor3000\relax}%
\providecommand \BibitemShut  [1]{\csname bibitem#1\endcsname}%
\let\auto@bib@innerbib\@empty
%</preamble>
\bibitem [{\citenamefont {Helled}\ \emph {et~al.}(2020)\citenamefont {Helled}, \citenamefont {Mazzola},\ and\ \citenamefont {Redmer}}]{helled2020understanding}%
  \BibitemOpen
  \bibfield  {author} {\bibinfo {author} {\bibfnamefont {R.}~\bibnamefont {Helled}}, \bibinfo {author} {\bibfnamefont {G.}~\bibnamefont {Mazzola}}, \ and\ \bibinfo {author} {\bibfnamefont {R.}~\bibnamefont {Redmer}},\ }\bibfield  {title} {\enquote {\bibinfo {title} {Understanding dense hydrogen at planetary conditions},}\ }\href@noop {} {\bibfield  {journal} {\bibinfo  {journal} {Nature Reviews Physics}\ }\textbf {\bibinfo {volume} {2}},\ \bibinfo {pages} {562} (\bibinfo {year} {2020})}\BibitemShut {NoStop}%
\bibitem [{\citenamefont {Stevenson}\ and\ \citenamefont {Salpeter}(1977{\natexlab{a}})}]{stevenson1977phase}%
  \BibitemOpen
  \bibfield  {author} {\bibinfo {author} {\bibfnamefont {D.~J.}\ \bibnamefont {Stevenson}}\ and\ \bibinfo {author} {\bibfnamefont {E.~E.}\ \bibnamefont {Salpeter}},\ }\bibfield  {title} {\enquote {\bibinfo {title} {The phase diagram and transport properties for hydrogen--helium fluid planets},}\ }\href {\doibase 10.1086/190478} {\bibfield  {journal} {\bibinfo  {journal} {The Astrophysical Journal Supplement Series}\ }\textbf {\bibinfo {volume} {35}},\ \bibinfo {pages} {221} (\bibinfo {year} {1977}{\natexlab{a}})}\BibitemShut {NoStop}%
\bibitem [{\citenamefont {Stevenson}\ and\ \citenamefont {Salpeter}(1977{\natexlab{b}})}]{stevenson1977dynamics}%
  \BibitemOpen
  \bibfield  {author} {\bibinfo {author} {\bibfnamefont {D.~J.}\ \bibnamefont {Stevenson}}\ and\ \bibinfo {author} {\bibfnamefont {E.~E.}\ \bibnamefont {Salpeter}},\ }\bibfield  {title} {\enquote {\bibinfo {title} {The dynamics and helium distribution in hydrogen--helium fluid planets},}\ }\href {\doibase 10.1086/190479} {\bibfield  {journal} {\bibinfo  {journal} {The Astrophysical Journal Supplement Series}\ }\textbf {\bibinfo {volume} {35}},\ \bibinfo {pages} {239} (\bibinfo {year} {1977}{\natexlab{b}})}\BibitemShut {NoStop}%
\bibitem [{\citenamefont {Helled}(2019)}]{Helled2019interiors}%
  \BibitemOpen
  \bibfield  {author} {\bibinfo {author} {\bibfnamefont {R.}~\bibnamefont {Helled}},\ }\href {\doibase 10.1093/acrefore/9780190647926.013.175} {\enquote {\bibinfo {title} {{The interiors of Jupiter and Saturn}},}\ }\bibinfo {howpublished} {Oxford Research Encyclopedia of Planetary Science} (\bibinfo {year} {2019})\BibitemShut {NoStop}%
\bibitem [{\citenamefont {McMahon}\ \emph {et~al.}(2012)\citenamefont {McMahon}, \citenamefont {Morales}, \citenamefont {Pierleoni},\ and\ \citenamefont {Ceperley}}]{mcmahon2012properties}%
  \BibitemOpen
  \bibfield  {author} {\bibinfo {author} {\bibfnamefont {J.~M.}\ \bibnamefont {McMahon}}, \bibinfo {author} {\bibfnamefont {M.~A.}\ \bibnamefont {Morales}}, \bibinfo {author} {\bibfnamefont {C.}~\bibnamefont {Pierleoni}}, \ and\ \bibinfo {author} {\bibfnamefont {D.~M.}\ \bibnamefont {Ceperley}},\ }\bibfield  {title} {\enquote {\bibinfo {title} {The properties of hydrogen and helium under extreme conditions},}\ }\href@noop {} {\bibfield  {journal} {\bibinfo  {journal} {Reviews of Modern Physics}\ }\textbf {\bibinfo {volume} {84}},\ \bibinfo {pages} {1607} (\bibinfo {year} {2012})}\BibitemShut {NoStop}%
\bibitem [{\citenamefont {Campbell}\ and\ \citenamefont {Synnott}(1985)}]{campbell1985gravity}%
  \BibitemOpen
  \bibfield  {author} {\bibinfo {author} {\bibfnamefont {J.}~\bibnamefont {Campbell}}\ and\ \bibinfo {author} {\bibfnamefont {S.}~\bibnamefont {Synnott}},\ }\bibfield  {title} {\enquote {\bibinfo {title} {{Gravity field of the Jovian system from Pioneer and Voyager tracking data}},}\ }\href@noop {} {\bibfield  {journal} {\bibinfo  {journal} {Astronomical Journal}\ }\textbf {\bibinfo {volume} {90}},\ \bibinfo {pages} {364} (\bibinfo {year} {1985})}\BibitemShut {NoStop}%
\bibitem [{\citenamefont {Bolton}\ \emph {et~al.}(2017)\citenamefont {Bolton}, \citenamefont {Adriani}, \citenamefont {Adumitroaie}, \citenamefont {Allison}, \citenamefont {Anderson}, \citenamefont {Atreya}, \citenamefont {Bloxham}, \citenamefont {Brown}, \citenamefont {Connerney}, \citenamefont {DeJong}, \citenamefont {Folkner}, \citenamefont {Gautier}, \citenamefont {Grassi}, \citenamefont {Gulkis}, \citenamefont {Guillot}, \citenamefont {Hansen}, \citenamefont {Hubbard}, \citenamefont {Iess}, \citenamefont {Ingersoll}, \citenamefont {Janssen}, \citenamefont {Jorgensen}, \citenamefont {Kaspi}, \citenamefont {Levin}, \citenamefont {Li}, \citenamefont {Lunine}, \citenamefont {Miguel}, \citenamefont {Mura}, \citenamefont {Orton}, \citenamefont {Owen}, \citenamefont {Ravine}, \citenamefont {Smith}, \citenamefont {Steffes}, \citenamefont {Stone}, \citenamefont {Stevenson}, \citenamefont {Thorne}, \citenamefont {Waite}, \citenamefont {Durante}, \citenamefont {Ebert}, \citenamefont {Greathouse}, \citenamefont
  {Hue}, \citenamefont {Parisi}, \citenamefont {Szalay},\ and\ \citenamefont {Wilson}}]{bolton2017jupiter}%
  \BibitemOpen
  \bibfield  {author} {\bibinfo {author} {\bibfnamefont {S.~J.}\ \bibnamefont {Bolton}}, \bibinfo {author} {\bibfnamefont {A.}~\bibnamefont {Adriani}}, \bibinfo {author} {\bibfnamefont {V.}~\bibnamefont {Adumitroaie}}, \bibinfo {author} {\bibfnamefont {M.}~\bibnamefont {Allison}}, \bibinfo {author} {\bibfnamefont {J.}~\bibnamefont {Anderson}}, \bibinfo {author} {\bibfnamefont {S.}~\bibnamefont {Atreya}}, \bibinfo {author} {\bibfnamefont {J.}~\bibnamefont {Bloxham}}, \bibinfo {author} {\bibfnamefont {S.}~\bibnamefont {Brown}}, \bibinfo {author} {\bibfnamefont {J.~E.~P.}\ \bibnamefont {Connerney}}, \bibinfo {author} {\bibfnamefont {E.}~\bibnamefont {DeJong}}, \bibinfo {author} {\bibfnamefont {W.}~\bibnamefont {Folkner}}, \bibinfo {author} {\bibfnamefont {D.}~\bibnamefont {Gautier}}, \bibinfo {author} {\bibfnamefont {D.}~\bibnamefont {Grassi}}, \bibinfo {author} {\bibfnamefont {S.}~\bibnamefont {Gulkis}}, \bibinfo {author} {\bibfnamefont {T.}~\bibnamefont {Guillot}}, \bibinfo {author} {\bibfnamefont
  {C.}~\bibnamefont {Hansen}}, \bibinfo {author} {\bibfnamefont {W.~B.}\ \bibnamefont {Hubbard}}, \bibinfo {author} {\bibfnamefont {L.}~\bibnamefont {Iess}}, \bibinfo {author} {\bibfnamefont {A.}~\bibnamefont {Ingersoll}}, \bibinfo {author} {\bibfnamefont {M.}~\bibnamefont {Janssen}}, \bibinfo {author} {\bibfnamefont {J.}~\bibnamefont {Jorgensen}}, \bibinfo {author} {\bibfnamefont {Y.}~\bibnamefont {Kaspi}}, \bibinfo {author} {\bibfnamefont {S.~M.}\ \bibnamefont {Levin}}, \bibinfo {author} {\bibfnamefont {C.}~\bibnamefont {Li}}, \bibinfo {author} {\bibfnamefont {J.}~\bibnamefont {Lunine}}, \bibinfo {author} {\bibfnamefont {Y.}~\bibnamefont {Miguel}}, \bibinfo {author} {\bibfnamefont {A.}~\bibnamefont {Mura}}, \bibinfo {author} {\bibfnamefont {G.}~\bibnamefont {Orton}}, \bibinfo {author} {\bibfnamefont {T.}~\bibnamefont {Owen}}, \bibinfo {author} {\bibfnamefont {M.}~\bibnamefont {Ravine}}, \bibinfo {author} {\bibfnamefont {E.}~\bibnamefont {Smith}}, \bibinfo {author} {\bibfnamefont {P.}~\bibnamefont
  {Steffes}}, \bibinfo {author} {\bibfnamefont {E.}~\bibnamefont {Stone}}, \bibinfo {author} {\bibfnamefont {D.}~\bibnamefont {Stevenson}}, \bibinfo {author} {\bibfnamefont {R.}~\bibnamefont {Thorne}}, \bibinfo {author} {\bibfnamefont {J.}~\bibnamefont {Waite}}, \bibinfo {author} {\bibfnamefont {D.}~\bibnamefont {Durante}}, \bibinfo {author} {\bibfnamefont {R.~W.}\ \bibnamefont {Ebert}}, \bibinfo {author} {\bibfnamefont {T.~K.}\ \bibnamefont {Greathouse}}, \bibinfo {author} {\bibfnamefont {V.}~\bibnamefont {Hue}}, \bibinfo {author} {\bibfnamefont {M.}~\bibnamefont {Parisi}}, \bibinfo {author} {\bibfnamefont {J.~R.}\ \bibnamefont {Szalay}}, \ and\ \bibinfo {author} {\bibfnamefont {R.}~\bibnamefont {Wilson}},\ }\bibfield  {title} {\enquote {\bibinfo {title} {{Jupiter’s interior and deep atmosphere: The initial pole-to-pole passes with the Juno spacecraft}},}\ }\href {\doibase 10.1126/science.aal2108} {\bibfield  {journal} {\bibinfo  {journal} {Science}\ }\textbf {\bibinfo {volume} {356}},\ \bibinfo {pages}
  {821} (\bibinfo {year} {2017})}\BibitemShut {NoStop}%
\bibitem [{\citenamefont {Spilker}(2019)}]{spilker2019cassini}%
  \BibitemOpen
  \bibfield  {author} {\bibinfo {author} {\bibfnamefont {L.}~\bibnamefont {Spilker}},\ }\bibfield  {title} {\enquote {\bibinfo {title} {{Cassini-Huygens’ exploration of the Saturn system: 13 years of discovery}},}\ }\href@noop {} {\bibfield  {journal} {\bibinfo  {journal} {Science}\ }\textbf {\bibinfo {volume} {364}},\ \bibinfo {pages} {1046} (\bibinfo {year} {2019})}\BibitemShut {NoStop}%
\bibitem [{\citenamefont {Young}(2003)}]{young2003galileo}%
  \BibitemOpen
  \bibfield  {author} {\bibinfo {author} {\bibfnamefont {R.~E.}\ \bibnamefont {Young}},\ }\bibfield  {title} {\enquote {\bibinfo {title} {{The Galileo probe: how it has changed our understanding of Jupiter}},}\ }\href@noop {} {\bibfield  {journal} {\bibinfo  {journal} {New Astronomy Reviews}\ }\textbf {\bibinfo {volume} {47}},\ \bibinfo {pages} {1} (\bibinfo {year} {2003})}\BibitemShut {NoStop}%
\bibitem [{\citenamefont {{von Zahn}}\ \emph {et~al.}(1998)\citenamefont {{von Zahn}}, \citenamefont {Hunten},\ and\ \citenamefont {Lehmacher}}]{von1998helium}%
  \BibitemOpen
  \bibfield  {author} {\bibinfo {author} {\bibfnamefont {U.}~\bibnamefont {{von Zahn}}}, \bibinfo {author} {\bibfnamefont {D.~M.}\ \bibnamefont {Hunten}}, \ and\ \bibinfo {author} {\bibfnamefont {G.}~\bibnamefont {Lehmacher}},\ }\bibfield  {title} {\enquote {\bibinfo {title} {{Helium in Jupiter's atmosphere: Results from the Galileo probe helium interferometer experiment}},}\ }\href@noop {} {\bibfield  {journal} {\bibinfo  {journal} {Journal of Geophysical Research: Planets}\ }\textbf {\bibinfo {volume} {103}},\ \bibinfo {pages} {22815} (\bibinfo {year} {1998})}\BibitemShut {NoStop}%
\bibitem [{\citenamefont {Fortney}\ \emph {et~al.}(2023)\citenamefont {Fortney}, \citenamefont {Militzer}, \citenamefont {Mankovich}, \citenamefont {Helled}, \citenamefont {Wahl}, \citenamefont {Nettelmann}, \citenamefont {Hubbard}, \citenamefont {Stevenson}, \citenamefont {Iess}, \citenamefont {Marley},\ and\ \citenamefont {Movshovitz}}]{fortney2023saturn}%
  \BibitemOpen
  \bibfield  {author} {\bibinfo {author} {\bibfnamefont {J.~J.}\ \bibnamefont {Fortney}}, \bibinfo {author} {\bibfnamefont {B.}~\bibnamefont {Militzer}}, \bibinfo {author} {\bibfnamefont {C.~R.}\ \bibnamefont {Mankovich}}, \bibinfo {author} {\bibfnamefont {R.}~\bibnamefont {Helled}}, \bibinfo {author} {\bibfnamefont {S.~M.}\ \bibnamefont {Wahl}}, \bibinfo {author} {\bibfnamefont {N.}~\bibnamefont {Nettelmann}}, \bibinfo {author} {\bibfnamefont {W.~B.}\ \bibnamefont {Hubbard}}, \bibinfo {author} {\bibfnamefont {D.~J.}\ \bibnamefont {Stevenson}}, \bibinfo {author} {\bibfnamefont {L.}~\bibnamefont {Iess}}, \bibinfo {author} {\bibfnamefont {M.~S.}\ \bibnamefont {Marley}}, \ and\ \bibinfo {author} {\bibfnamefont {N.}~\bibnamefont {Movshovitz}},\ }\href {\doibase 10.48550/arXiv.2304.09215} {\enquote {\bibinfo {title} {{Saturn's interior after the Cassini Grand Finale}},}\ } (\bibinfo {year} {2023}),\ \Eprint {http://arxiv.org/abs/2304.09215} {arXiv:2304.09215 [astro-ph.EP]} \BibitemShut {NoStop}%
\bibitem [{\citenamefont {Proffitt}(1994)}]{Proffitt1994proto}%
  \BibitemOpen
  \bibfield  {author} {\bibinfo {author} {\bibfnamefont {C.~R.}\ \bibnamefont {Proffitt}},\ }\bibfield  {title} {\enquote {\bibinfo {title} {Effects of heavy-element settling on solar-neutrino fluxes and interior structure},}\ }\href {\doibase 10.1086/174030} {\bibfield  {journal} {\bibinfo  {journal} {The Astrophysical Journal}\ }\textbf {\bibinfo {volume} {425}},\ \bibinfo {pages} {849} (\bibinfo {year} {1994})}\BibitemShut {NoStop}%
\bibitem [{\citenamefont {Bahcall}\ \emph {et~al.}(1995)\citenamefont {Bahcall}, \citenamefont {Pinsonneault},\ and\ \citenamefont {Wasserburg}}]{bahcall1995solar}%
  \BibitemOpen
  \bibfield  {author} {\bibinfo {author} {\bibfnamefont {J.~N.}\ \bibnamefont {Bahcall}}, \bibinfo {author} {\bibfnamefont {M.}~\bibnamefont {Pinsonneault}}, \ and\ \bibinfo {author} {\bibfnamefont {G.}~\bibnamefont {Wasserburg}},\ }\bibfield  {title} {\enquote {\bibinfo {title} {Solar models with helium and heavy-element diffusion},}\ }\href {\doibase 10.1103/RevModPhys.67.781} {\bibfield  {journal} {\bibinfo  {journal} {Reviews of Modern Physics}\ }\textbf {\bibinfo {volume} {67}},\ \bibinfo {pages} {781} (\bibinfo {year} {1995})}\BibitemShut {NoStop}%
\bibitem [{\citenamefont {Low}(1966)}]{low1966observations}%
  \BibitemOpen
  \bibfield  {author} {\bibinfo {author} {\bibfnamefont {F.~J.}\ \bibnamefont {Low}},\ }\bibfield  {title} {\enquote {\bibinfo {title} {{Observations of Venus, Jupiter, and Saturn at $\lambda$20 $\mu$.}}}\ }\href {\doibase 10.1086/110110} {\bibfield  {journal} {\bibinfo  {journal} {The Astronomical Journal}\ }\textbf {\bibinfo {volume} {71}},\ \bibinfo {pages} {391} (\bibinfo {year} {1966})}\BibitemShut {NoStop}%
\bibitem [{\citenamefont {Militzer}\ \emph {et~al.}(2016)\citenamefont {Militzer}, \citenamefont {Soubiran}, \citenamefont {Wahl},\ and\ \citenamefont {Hubbard}}]{militzer2016understanding}%
  \BibitemOpen
  \bibfield  {author} {\bibinfo {author} {\bibfnamefont {B.}~\bibnamefont {Militzer}}, \bibinfo {author} {\bibfnamefont {F.}~\bibnamefont {Soubiran}}, \bibinfo {author} {\bibfnamefont {S.~M.}\ \bibnamefont {Wahl}}, \ and\ \bibinfo {author} {\bibfnamefont {W.}~\bibnamefont {Hubbard}},\ }\bibfield  {title} {\enquote {\bibinfo {title} {{Understanding Jupiter's interior}},}\ }\href@noop {} {\bibfield  {journal} {\bibinfo  {journal} {Journal of Geophysical Research: Planets}\ }\textbf {\bibinfo {volume} {121}},\ \bibinfo {pages} {1552} (\bibinfo {year} {2016})}\BibitemShut {NoStop}%
\bibitem [{\citenamefont {Militzer}\ and\ \citenamefont {Hubbard}(2024)}]{militzer2024study}%
  \BibitemOpen
  \bibfield  {author} {\bibinfo {author} {\bibfnamefont {B.}~\bibnamefont {Militzer}}\ and\ \bibinfo {author} {\bibfnamefont {W.~B.}\ \bibnamefont {Hubbard}},\ }\bibfield  {title} {\enquote {\bibinfo {title} {{Study of Jupiter’s interior: Comparison of 2, 3, 4, 5, and 6 layer models}},}\ }\href@noop {} {\bibfield  {journal} {\bibinfo  {journal} {Icarus}\ }\textbf {\bibinfo {volume} {411}},\ \bibinfo {pages} {115955} (\bibinfo {year} {2024})}\BibitemShut {NoStop}%
\bibitem [{\citenamefont {Duarte}\ \emph {et~al.}(2018)\citenamefont {Duarte}, \citenamefont {Wicht},\ and\ \citenamefont {Gastine}}]{duarte2018physical}%
  \BibitemOpen
  \bibfield  {author} {\bibinfo {author} {\bibfnamefont {L.~D.}\ \bibnamefont {Duarte}}, \bibinfo {author} {\bibfnamefont {J.}~\bibnamefont {Wicht}}, \ and\ \bibinfo {author} {\bibfnamefont {T.}~\bibnamefont {Gastine}},\ }\bibfield  {title} {\enquote {\bibinfo {title} {{Physical conditions for Jupiter-like dynamo models}},}\ }\href {\doibase 10.1016/j.icarus.2017.07.016} {\bibfield  {journal} {\bibinfo  {journal} {Icarus}\ }\textbf {\bibinfo {volume} {299}},\ \bibinfo {pages} {206} (\bibinfo {year} {2018})}\BibitemShut {NoStop}%
\bibitem [{\citenamefont {Howard}\ \emph {et~al.}(2024)\citenamefont {Howard}, \citenamefont {M{\"u}ller},\ and\ \citenamefont {Helled}}]{howard2024evolution}%
  \BibitemOpen
  \bibfield  {author} {\bibinfo {author} {\bibfnamefont {S.}~\bibnamefont {Howard}}, \bibinfo {author} {\bibfnamefont {S.}~\bibnamefont {M{\"u}ller}}, \ and\ \bibinfo {author} {\bibfnamefont {R.}~\bibnamefont {Helled}},\ }\bibfield  {title} {\enquote {\bibinfo {title} {{Evolution of Jupiter and Saturn with helium rain}},}\ }\href@noop {} {\bibfield  {journal} {\bibinfo  {journal} {Astronomy \& Astrophysics}\ }\textbf {\bibinfo {volume} {689}},\ \bibinfo {pages} {A15} (\bibinfo {year} {2024})}\BibitemShut {NoStop}%
\bibitem [{\citenamefont {Brygoo}\ \emph {et~al.}(2021)\citenamefont {Brygoo}, \citenamefont {Loubeyre}, \citenamefont {Millot}, \citenamefont {Rygg}, \citenamefont {Celliers}, \citenamefont {Eggert}, \citenamefont {Jeanloz},\ and\ \citenamefont {Collins}}]{brygoo2021evidence}%
  \BibitemOpen
  \bibfield  {author} {\bibinfo {author} {\bibfnamefont {S.}~\bibnamefont {Brygoo}}, \bibinfo {author} {\bibfnamefont {P.}~\bibnamefont {Loubeyre}}, \bibinfo {author} {\bibfnamefont {M.}~\bibnamefont {Millot}}, \bibinfo {author} {\bibfnamefont {J.}~\bibnamefont {Rygg}}, \bibinfo {author} {\bibfnamefont {P.}~\bibnamefont {Celliers}}, \bibinfo {author} {\bibfnamefont {J.}~\bibnamefont {Eggert}}, \bibinfo {author} {\bibfnamefont {R.}~\bibnamefont {Jeanloz}}, \ and\ \bibinfo {author} {\bibfnamefont {G.}~\bibnamefont {Collins}},\ }\bibfield  {title} {\enquote {\bibinfo {title} {{Evidence of hydrogen-helium immiscibility at Jupiter-interior conditions}},}\ }\href@noop {} {\bibfield  {journal} {\bibinfo  {journal} {Nature}\ }\textbf {\bibinfo {volume} {593}},\ \bibinfo {pages} {517} (\bibinfo {year} {2021})}\BibitemShut {NoStop}%
\bibitem [{\citenamefont {Lorenzen}\ \emph {et~al.}(2009)\citenamefont {Lorenzen}, \citenamefont {Holst},\ and\ \citenamefont {Redmer}}]{lorenzen2009demixing}%
  \BibitemOpen
  \bibfield  {author} {\bibinfo {author} {\bibfnamefont {W.}~\bibnamefont {Lorenzen}}, \bibinfo {author} {\bibfnamefont {B.}~\bibnamefont {Holst}}, \ and\ \bibinfo {author} {\bibfnamefont {R.}~\bibnamefont {Redmer}},\ }\bibfield  {title} {\enquote {\bibinfo {title} {Demixing of hydrogen and helium at megabar pressures},}\ }\href@noop {} {\bibfield  {journal} {\bibinfo  {journal} {Physical Review Letters}\ }\textbf {\bibinfo {volume} {102}},\ \bibinfo {pages} {115701} (\bibinfo {year} {2009})}\BibitemShut {NoStop}%
\bibitem [{\citenamefont {Morales}\ \emph {et~al.}(2009)\citenamefont {Morales}, \citenamefont {Schwegler}, \citenamefont {Ceperley}, \citenamefont {Pierleoni}, \citenamefont {Hamel},\ and\ \citenamefont {Caspersen}}]{morales2009}%
  \BibitemOpen
  \bibfield  {author} {\bibinfo {author} {\bibfnamefont {M.~A.}\ \bibnamefont {Morales}}, \bibinfo {author} {\bibfnamefont {E.}~\bibnamefont {Schwegler}}, \bibinfo {author} {\bibfnamefont {D.}~\bibnamefont {Ceperley}}, \bibinfo {author} {\bibfnamefont {C.}~\bibnamefont {Pierleoni}}, \bibinfo {author} {\bibfnamefont {S.}~\bibnamefont {Hamel}}, \ and\ \bibinfo {author} {\bibfnamefont {K.}~\bibnamefont {Caspersen}},\ }\bibfield  {title} {\enquote {\bibinfo {title} {Phase separation in hydrogen–helium mixtures at mbar pressures},}\ }\href {\doibase 10.1073/pnas.0812581106} {\bibfield  {journal} {\bibinfo  {journal} {Proceedings of the National Academy of Sciences}\ }\textbf {\bibinfo {volume} {106}},\ \bibinfo {pages} {1324} (\bibinfo {year} {2009})}\BibitemShut {NoStop}%
\bibitem [{\citenamefont {Lorenzen}\ \emph {et~al.}(2011)\citenamefont {Lorenzen}, \citenamefont {Holst},\ and\ \citenamefont {Redmer}}]{lorenzen2011metallization}%
  \BibitemOpen
  \bibfield  {author} {\bibinfo {author} {\bibfnamefont {W.}~\bibnamefont {Lorenzen}}, \bibinfo {author} {\bibfnamefont {B.}~\bibnamefont {Holst}}, \ and\ \bibinfo {author} {\bibfnamefont {R.}~\bibnamefont {Redmer}},\ }\bibfield  {title} {\enquote {\bibinfo {title} {Metallization in hydrogen-helium mixtures},}\ }\href {\doibase 10.1103/PhysRevB.84.235109} {\bibfield  {journal} {\bibinfo  {journal} {Physical Review B}\ }\textbf {\bibinfo {volume} {84}},\ \bibinfo {pages} {235109} (\bibinfo {year} {2011})}\BibitemShut {NoStop}%
\bibitem [{\citenamefont {Morales}\ \emph {et~al.}(2013{\natexlab{a}})\citenamefont {Morales}, \citenamefont {Hamel}, \citenamefont {Caspersen},\ and\ \citenamefont {Schwegler}}]{morales2013hydrogen}%
  \BibitemOpen
  \bibfield  {author} {\bibinfo {author} {\bibfnamefont {M.~A.}\ \bibnamefont {Morales}}, \bibinfo {author} {\bibfnamefont {S.}~\bibnamefont {Hamel}}, \bibinfo {author} {\bibfnamefont {K.}~\bibnamefont {Caspersen}}, \ and\ \bibinfo {author} {\bibfnamefont {E.}~\bibnamefont {Schwegler}},\ }\bibfield  {title} {\enquote {\bibinfo {title} {Hydrogen-helium demixing from first principles: From diamond anvil cells to planetary interiors},}\ }\href {\doibase 10.1103/PhysRevB.87.174105} {\bibfield  {journal} {\bibinfo  {journal} {Physical Review B}\ }\textbf {\bibinfo {volume} {87}},\ \bibinfo {pages} {174105} (\bibinfo {year} {2013}{\natexlab{a}})}\BibitemShut {NoStop}%
\bibitem [{\citenamefont {Sch{\"o}ttler}\ and\ \citenamefont {Redmer}(2018)}]{schottler2018ab}%
  \BibitemOpen
  \bibfield  {author} {\bibinfo {author} {\bibfnamefont {M.}~\bibnamefont {Sch{\"o}ttler}}\ and\ \bibinfo {author} {\bibfnamefont {R.}~\bibnamefont {Redmer}},\ }\bibfield  {title} {\enquote {\bibinfo {title} {{\it Ab initio} calculation of the miscibility diagram for hydrogen-helium mixtures},}\ }\href@noop {} {\bibfield  {journal} {\bibinfo  {journal} {Physical Review Letters}\ }\textbf {\bibinfo {volume} {120}},\ \bibinfo {pages} {115703} (\bibinfo {year} {2018})}\BibitemShut {NoStop}%
\bibitem [{\citenamefont {Chang}\ \emph {et~al.}(2024)\citenamefont {Chang}, \citenamefont {Chen}, \citenamefont {Zeng}, \citenamefont {Wang}, \citenamefont {Chen}, \citenamefont {Tong}, \citenamefont {Yu}, \citenamefont {Kang}, \citenamefont {Zhang}, \citenamefont {Guo}, \citenamefont {Hou}, \citenamefont {Zhao}, \citenamefont {Yao}, \citenamefont {Ma},\ and\ \citenamefont {Dai}}]{chang2024theoretical}%
  \BibitemOpen
  \bibfield  {author} {\bibinfo {author} {\bibfnamefont {X.}~\bibnamefont {Chang}}, \bibinfo {author} {\bibfnamefont {B.}~\bibnamefont {Chen}}, \bibinfo {author} {\bibfnamefont {Q.}~\bibnamefont {Zeng}}, \bibinfo {author} {\bibfnamefont {H.}~\bibnamefont {Wang}}, \bibinfo {author} {\bibfnamefont {K.}~\bibnamefont {Chen}}, \bibinfo {author} {\bibfnamefont {Q.}~\bibnamefont {Tong}}, \bibinfo {author} {\bibfnamefont {X.}~\bibnamefont {Yu}}, \bibinfo {author} {\bibfnamefont {D.}~\bibnamefont {Kang}}, \bibinfo {author} {\bibfnamefont {S.}~\bibnamefont {Zhang}}, \bibinfo {author} {\bibfnamefont {F.}~\bibnamefont {Guo}}, \bibinfo {author} {\bibfnamefont {Y.}~\bibnamefont {Hou}}, \bibinfo {author} {\bibfnamefont {Z.}~\bibnamefont {Zhao}}, \bibinfo {author} {\bibfnamefont {Y.}~\bibnamefont {Yao}}, \bibinfo {author} {\bibfnamefont {Y.}~\bibnamefont {Ma}}, \ and\ \bibinfo {author} {\bibfnamefont {J.}~\bibnamefont {Dai}},\ }\bibfield  {title} {\enquote {\bibinfo {title} {{Theoretical evidence of H-He demixing under
  Jupiter and Saturn conditions}},}\ }\href {\doibase 10.1038/s41467-024-52868-4} {\bibfield  {journal} {\bibinfo  {journal} {Nature Communications}\ }\textbf {\bibinfo {volume} {15}},\ \bibinfo {pages} {8543} (\bibinfo {year} {2024})}\BibitemShut {NoStop}%
\bibitem [{\citenamefont {Mori-S{\'a}nchez}\ \emph {et~al.}(2008)\citenamefont {Mori-S{\'a}nchez}, \citenamefont {Cohen},\ and\ \citenamefont {Yang}}]{mori2008localization}%
  \BibitemOpen
  \bibfield  {author} {\bibinfo {author} {\bibfnamefont {P.}~\bibnamefont {Mori-S{\'a}nchez}}, \bibinfo {author} {\bibfnamefont {A.~J.}\ \bibnamefont {Cohen}}, \ and\ \bibinfo {author} {\bibfnamefont {W.}~\bibnamefont {Yang}},\ }\bibfield  {title} {\enquote {\bibinfo {title} {Localization and delocalization errors in density functional theory and implications for band-gap prediction},}\ }\href@noop {} {\bibfield  {journal} {\bibinfo  {journal} {Physical Review Letters}\ }\textbf {\bibinfo {volume} {100}},\ \bibinfo {pages} {146401} (\bibinfo {year} {2008})}\BibitemShut {NoStop}%
\bibitem [{\citenamefont {Cohen}\ \emph {et~al.}(2012)\citenamefont {Cohen}, \citenamefont {Mori-S{\'a}nchez},\ and\ \citenamefont {Yang}}]{cohen2012challenges}%
  \BibitemOpen
  \bibfield  {author} {\bibinfo {author} {\bibfnamefont {A.~J.}\ \bibnamefont {Cohen}}, \bibinfo {author} {\bibfnamefont {P.}~\bibnamefont {Mori-S{\'a}nchez}}, \ and\ \bibinfo {author} {\bibfnamefont {W.}~\bibnamefont {Yang}},\ }\bibfield  {title} {\enquote {\bibinfo {title} {Challenges for density functional theory},}\ }\href@noop {} {\bibfield  {journal} {\bibinfo  {journal} {Chemical Reviews}\ }\textbf {\bibinfo {volume} {112}},\ \bibinfo {pages} {289} (\bibinfo {year} {2012})}\BibitemShut {NoStop}%
\bibitem [{\citenamefont {Burke}(2012)}]{burke2012perspective}%
  \BibitemOpen
  \bibfield  {author} {\bibinfo {author} {\bibfnamefont {K.}~\bibnamefont {Burke}},\ }\bibfield  {title} {\enquote {\bibinfo {title} {Perspective on density functional theory},}\ }\href {\doibase 10.1063/1.4704546} {\bibfield  {journal} {\bibinfo  {journal} {The Journal of Chemical Physics}\ }\textbf {\bibinfo {volume} {136}},\ \bibinfo {pages} {150901} (\bibinfo {year} {2012})}\BibitemShut {NoStop}%
\bibitem [{\citenamefont {Sim}\ \emph {et~al.}(2022)\citenamefont {Sim}, \citenamefont {Song}, \citenamefont {Vuckovic},\ and\ \citenamefont {Burke}}]{sim2022improving}%
  \BibitemOpen
  \bibfield  {author} {\bibinfo {author} {\bibfnamefont {E.}~\bibnamefont {Sim}}, \bibinfo {author} {\bibfnamefont {S.}~\bibnamefont {Song}}, \bibinfo {author} {\bibfnamefont {S.}~\bibnamefont {Vuckovic}}, \ and\ \bibinfo {author} {\bibfnamefont {K.}~\bibnamefont {Burke}},\ }\bibfield  {title} {\enquote {\bibinfo {title} {Improving results by improving densities: Density-corrected density functional theory},}\ }\href@noop {} {\bibfield  {journal} {\bibinfo  {journal} {Journal of the American Chemical Society}\ }\textbf {\bibinfo {volume} {144}},\ \bibinfo {pages} {6625} (\bibinfo {year} {2022})}\BibitemShut {NoStop}%
\bibitem [{\citenamefont {Cozza}\ \emph {et~al.}(2026)\citenamefont {Cozza}, \citenamefont {Nakano}, \citenamefont {Howard}, \citenamefont {Xie}, \citenamefont {Helled},\ and\ \citenamefont {Mazzola}}]{cozza2026denser}%
  \BibitemOpen
  \bibfield  {author} {\bibinfo {author} {\bibfnamefont {C.}~\bibnamefont {Cozza}}, \bibinfo {author} {\bibfnamefont {K.}~\bibnamefont {Nakano}}, \bibinfo {author} {\bibfnamefont {S.}~\bibnamefont {Howard}}, \bibinfo {author} {\bibfnamefont {H.}~\bibnamefont {Xie}}, \bibinfo {author} {\bibfnamefont {R.}~\bibnamefont {Helled}}, \ and\ \bibinfo {author} {\bibfnamefont {G.}~\bibnamefont {Mazzola}},\ }\bibfield  {title} {\enquote {\bibinfo {title} {Denser hydrogen inferred from first-principles simulations challenges jupiter’s interior models},}\ }\href@noop {} {\bibfield  {journal} {\bibinfo  {journal} {Physical Review Research}\ }\textbf {\bibinfo {volume} {8}},\ \bibinfo {pages} {013089} (\bibinfo {year} {2026})}\BibitemShut {NoStop}%
\bibitem [{\citenamefont {Clay}\ \emph {et~al.}(2016)\citenamefont {Clay}, \citenamefont {Holzmann}, \citenamefont {Ceperley},\ and\ \citenamefont {Morales}}]{clay2016benchmarking}%
  \BibitemOpen
  \bibfield  {author} {\bibinfo {author} {\bibfnamefont {R.~C.}\ \bibnamefont {Clay}, \bibfnamefont {III}}, \bibinfo {author} {\bibfnamefont {M.}~\bibnamefont {Holzmann}}, \bibinfo {author} {\bibfnamefont {D.~M.}\ \bibnamefont {Ceperley}}, \ and\ \bibinfo {author} {\bibfnamefont {M.~A.}\ \bibnamefont {Morales}},\ }\bibfield  {title} {\enquote {\bibinfo {title} {Benchmarking density functionals for hydrogen-helium mixtures with quantum {Monte Carlo}: Energetics, pressures, and forces},}\ }\href {\doibase 10.1103/PhysRevB.93.035121} {\bibfield  {journal} {\bibinfo  {journal} {Physical Review B}\ }\textbf {\bibinfo {volume} {93}},\ \bibinfo {pages} {035121} (\bibinfo {year} {2016})}\BibitemShut {NoStop}%
\bibitem [{\citenamefont {Cheng}\ \emph {et~al.}(2020)\citenamefont {Cheng}, \citenamefont {Mazzola}, \citenamefont {Pickard},\ and\ \citenamefont {Ceriotti}}]{cheng2020evidence}%
  \BibitemOpen
  \bibfield  {author} {\bibinfo {author} {\bibfnamefont {B.}~\bibnamefont {Cheng}}, \bibinfo {author} {\bibfnamefont {G.}~\bibnamefont {Mazzola}}, \bibinfo {author} {\bibfnamefont {C.~J.}\ \bibnamefont {Pickard}}, \ and\ \bibinfo {author} {\bibfnamefont {M.}~\bibnamefont {Ceriotti}},\ }\bibfield  {title} {\enquote {\bibinfo {title} {Evidence for supercritical behaviour of high-pressure liquid hydrogen},}\ }\href@noop {} {\bibfield  {journal} {\bibinfo  {journal} {Nature}\ }\textbf {\bibinfo {volume} {585}},\ \bibinfo {pages} {217} (\bibinfo {year} {2020})}\BibitemShut {NoStop}%
\bibitem [{\citenamefont {Cheng}\ \emph {et~al.}(2021)\citenamefont {Cheng}, \citenamefont {Mazzola}, \citenamefont {Pickard},\ and\ \citenamefont {Ceriotti}}]{cheng2021reply}%
  \BibitemOpen
  \bibfield  {author} {\bibinfo {author} {\bibfnamefont {B.}~\bibnamefont {Cheng}}, \bibinfo {author} {\bibfnamefont {G.}~\bibnamefont {Mazzola}}, \bibinfo {author} {\bibfnamefont {C.~J.}\ \bibnamefont {Pickard}}, \ and\ \bibinfo {author} {\bibfnamefont {M.}~\bibnamefont {Ceriotti}},\ }\bibfield  {title} {\enquote {\bibinfo {title} {Reply to: On the liquid--liquid phase transition of dense hydrogen},}\ }\href@noop {} {\bibfield  {journal} {\bibinfo  {journal} {Nature}\ }\textbf {\bibinfo {volume} {600}},\ \bibinfo {pages} {E15} (\bibinfo {year} {2021})}\BibitemShut {NoStop}%
\bibitem [{\citenamefont {Deringer}\ \emph {et~al.}(2019)\citenamefont {Deringer}, \citenamefont {Caro},\ and\ \citenamefont {Cs{\'a}nyi}}]{Deringer2019}%
  \BibitemOpen
  \bibfield  {author} {\bibinfo {author} {\bibfnamefont {V.~L.}\ \bibnamefont {Deringer}}, \bibinfo {author} {\bibfnamefont {M.~A.}\ \bibnamefont {Caro}}, \ and\ \bibinfo {author} {\bibfnamefont {G.}~\bibnamefont {Cs{\'a}nyi}},\ }\bibfield  {title} {\enquote {\bibinfo {title} {Machine learning interatomic potentials as emerging tools for materials science},}\ }\href {\doibase 10.1002/adma.201902765} {\bibfield  {journal} {\bibinfo  {journal} {Advanced Materials}\ }\textbf {\bibinfo {volume} {31}},\ \bibinfo {pages} {1902765} (\bibinfo {year} {2019})}\BibitemShut {NoStop}%
\bibitem [{\citenamefont {Perdew}\ \emph {et~al.}(1996)\citenamefont {Perdew}, \citenamefont {Burke},\ and\ \citenamefont {Ernzerhof}}]{perdew1996generalized}%
  \BibitemOpen
  \bibfield  {author} {\bibinfo {author} {\bibfnamefont {J.~P.}\ \bibnamefont {Perdew}}, \bibinfo {author} {\bibfnamefont {K.}~\bibnamefont {Burke}}, \ and\ \bibinfo {author} {\bibfnamefont {M.}~\bibnamefont {Ernzerhof}},\ }\bibfield  {title} {\enquote {\bibinfo {title} {Generalized gradient approximation made simple},}\ }\href {\doibase 10.1103/PhysRevLett.77.3865} {\bibfield  {journal} {\bibinfo  {journal} {Physical Review Letters}\ }\textbf {\bibinfo {volume} {77}},\ \bibinfo {pages} {3865} (\bibinfo {year} {1996})}\BibitemShut {NoStop}%
\bibitem [{\citenamefont {Lee}\ \emph {et~al.}(2010)\citenamefont {Lee}, \citenamefont {Murray}, \citenamefont {Kong}, \citenamefont {Lundqvist},\ and\ \citenamefont {Langreth}}]{lee2010higher}%
  \BibitemOpen
  \bibfield  {author} {\bibinfo {author} {\bibfnamefont {K.}~\bibnamefont {Lee}}, \bibinfo {author} {\bibfnamefont {{\'E}.~D.}\ \bibnamefont {Murray}}, \bibinfo {author} {\bibfnamefont {L.}~\bibnamefont {Kong}}, \bibinfo {author} {\bibfnamefont {B.~I.}\ \bibnamefont {Lundqvist}}, \ and\ \bibinfo {author} {\bibfnamefont {D.~C.}\ \bibnamefont {Langreth}},\ }\bibfield  {title} {\enquote {\bibinfo {title} {Higher-accuracy van der {Waals} density functional},}\ }\href {\doibase 10.1103/PhysRevB.82.081101} {\bibfield  {journal} {\bibinfo  {journal} {Physical Review B}\ }\textbf {\bibinfo {volume} {82}},\ \bibinfo {pages} {081101} (\bibinfo {year} {2010})}\BibitemShut {NoStop}%
\bibitem [{\citenamefont {Heyd}\ \emph {et~al.}(2003)\citenamefont {Heyd}, \citenamefont {Scuseria},\ and\ \citenamefont {Ernzerhof}}]{heyd2003hybrid}%
  \BibitemOpen
  \bibfield  {author} {\bibinfo {author} {\bibfnamefont {J.}~\bibnamefont {Heyd}}, \bibinfo {author} {\bibfnamefont {G.~E.}\ \bibnamefont {Scuseria}}, \ and\ \bibinfo {author} {\bibfnamefont {M.}~\bibnamefont {Ernzerhof}},\ }\bibfield  {title} {\enquote {\bibinfo {title} {Hybrid functionals based on a screened {Coulomb} potential},}\ }\href@noop {} {\bibfield  {journal} {\bibinfo  {journal} {The Journal of Chemical Physics}\ }\textbf {\bibinfo {volume} {118}},\ \bibinfo {pages} {8207} (\bibinfo {year} {2003})}\BibitemShut {NoStop}%
\bibitem [{\citenamefont {Heyd}\ \emph {et~al.}(2006)\citenamefont {Heyd}, \citenamefont {Scuseria},\ and\ \citenamefont {Ernzerhof}}]{heyd2006hybrid}%
  \BibitemOpen
  \bibfield  {author} {\bibinfo {author} {\bibfnamefont {J.}~\bibnamefont {Heyd}}, \bibinfo {author} {\bibfnamefont {G.~E.}\ \bibnamefont {Scuseria}}, \ and\ \bibinfo {author} {\bibfnamefont {M.}~\bibnamefont {Ernzerhof}},\ }\bibfield  {title} {\enquote {\bibinfo {title} {{Erratum: “Hybrid functionals based on a screened Coulomb potential” }},}\ }\href {\doibase 10.1063/1.2204597} {\bibfield  {journal} {\bibinfo  {journal} {The Journal of Chemical Physics}\ }\textbf {\bibinfo {volume} {124}},\ \bibinfo {pages} {219906} (\bibinfo {year} {2006})}\BibitemShut {NoStop}%
\bibitem [{\citenamefont {Cheng}(2022)}]{cheng2022computing}%
  \BibitemOpen
  \bibfield  {author} {\bibinfo {author} {\bibfnamefont {B.}~\bibnamefont {Cheng}},\ }\bibfield  {title} {\enquote {\bibinfo {title} {Computing chemical potentials of solutions from structure factors},}\ }\href {\doibase 10.1063/5.0107059} {\bibfield  {journal} {\bibinfo  {journal} {The Journal of Chemical Physics}\ }\textbf {\bibinfo {volume} {157}},\ \bibinfo {pages} {121101} (\bibinfo {year} {2022})}\BibitemShut {NoStop}%
\bibitem [{\citenamefont {Redlich}\ and\ \citenamefont {Kister}(1948)}]{redlich1948thermodynamics}%
  \BibitemOpen
  \bibfield  {author} {\bibinfo {author} {\bibfnamefont {O.}~\bibnamefont {Redlich}}\ and\ \bibinfo {author} {\bibfnamefont {A.}~\bibnamefont {Kister}},\ }\bibfield  {title} {\enquote {\bibinfo {title} {Thermodynamics of nonelectrolyte solutions - {\it x-y-t} relations in a binary system},}\ }\href {\doibase 10.1021/ie50458a035} {\bibfield  {journal} {\bibinfo  {journal} {Industrial \& Engineering Chemistry}\ }\textbf {\bibinfo {volume} {40}},\ \bibinfo {pages} {341} (\bibinfo {year} {1948})}\BibitemShut {NoStop}%
\bibitem [{\citenamefont {Militzer}\ and\ \citenamefont {Hubbard}(2013)}]{militzer2013ab}%
  \BibitemOpen
  \bibfield  {author} {\bibinfo {author} {\bibfnamefont {B.}~\bibnamefont {Militzer}}\ and\ \bibinfo {author} {\bibfnamefont {W.~B.}\ \bibnamefont {Hubbard}},\ }\bibfield  {title} {\enquote {\bibinfo {title} {{\it Ab initio} equation of state for hydrogen--helium mixtures with recalibration of the giant-planet mass--radius relation},}\ }\href@noop {} {\bibfield  {journal} {\bibinfo  {journal} {The Astrophysical Journal}\ }\textbf {\bibinfo {volume} {774}},\ \bibinfo {pages} {148} (\bibinfo {year} {2013})}\BibitemShut {NoStop}%
\bibitem [{\citenamefont {Nettelmann}\ \emph {et~al.}(2008)\citenamefont {Nettelmann}, \citenamefont {Holst}, \citenamefont {Kietzmann}, \citenamefont {French}, \citenamefont {Redmer},\ and\ \citenamefont {Blaschke}}]{nettelmann2008ab}%
  \BibitemOpen
  \bibfield  {author} {\bibinfo {author} {\bibfnamefont {N.}~\bibnamefont {Nettelmann}}, \bibinfo {author} {\bibfnamefont {B.}~\bibnamefont {Holst}}, \bibinfo {author} {\bibfnamefont {A.}~\bibnamefont {Kietzmann}}, \bibinfo {author} {\bibfnamefont {M.}~\bibnamefont {French}}, \bibinfo {author} {\bibfnamefont {R.}~\bibnamefont {Redmer}}, \ and\ \bibinfo {author} {\bibfnamefont {D.}~\bibnamefont {Blaschke}},\ }\bibfield  {title} {\enquote {\bibinfo {title} {{{\it Ab initio} equation of state data for hydrogen, helium, and water and the internal structure of Jupiter}},}\ }\href {\doibase 10.1086/589806} {\bibfield  {journal} {\bibinfo  {journal} {The Astrophysical Journal}\ }\textbf {\bibinfo {volume} {683}},\ \bibinfo {pages} {1217} (\bibinfo {year} {2008})}\BibitemShut {NoStop}%
\bibitem [{\citenamefont {Nettelmann}\ \emph {et~al.}(2013)\citenamefont {Nettelmann}, \citenamefont {P{\"u}stow},\ and\ \citenamefont {Redmer}}]{nettelmann2013saturn}%
  \BibitemOpen
  \bibfield  {author} {\bibinfo {author} {\bibfnamefont {N.}~\bibnamefont {Nettelmann}}, \bibinfo {author} {\bibfnamefont {R.}~\bibnamefont {P{\"u}stow}}, \ and\ \bibinfo {author} {\bibfnamefont {R.}~\bibnamefont {Redmer}},\ }\bibfield  {title} {\enquote {\bibinfo {title} {{Saturn layered structure and homogeneous evolution models with different EOSs}},}\ }\href@noop {} {\bibfield  {journal} {\bibinfo  {journal} {Icarus}\ }\textbf {\bibinfo {volume} {225}},\ \bibinfo {pages} {548} (\bibinfo {year} {2013})}\BibitemShut {NoStop}%
\bibitem [{\citenamefont {Mankovich}\ and\ \citenamefont {Fortney}(2020)}]{mankovich2020evidence}%
  \BibitemOpen
  \bibfield  {author} {\bibinfo {author} {\bibfnamefont {C.~R.}\ \bibnamefont {Mankovich}}\ and\ \bibinfo {author} {\bibfnamefont {J.~J.}\ \bibnamefont {Fortney}},\ }\bibfield  {title} {\enquote {\bibinfo {title} {{Evidence for a dichotomy in the interior structures of Jupiter and Saturn from helium phase separation}},}\ }\href@noop {} {\bibfield  {journal} {\bibinfo  {journal} {The Astrophysical Journal}\ }\textbf {\bibinfo {volume} {889}},\ \bibinfo {pages} {51} (\bibinfo {year} {2020})}\BibitemShut {NoStop}%
\bibitem [{\citenamefont {Singraber}\ \emph {et~al.}(2019)\citenamefont {Singraber}, \citenamefont {Morawietz}, \citenamefont {Behler},\ and\ \citenamefont {Dellago}}]{singraber2019parallel}%
  \BibitemOpen
  \bibfield  {author} {\bibinfo {author} {\bibfnamefont {A.}~\bibnamefont {Singraber}}, \bibinfo {author} {\bibfnamefont {T.}~\bibnamefont {Morawietz}}, \bibinfo {author} {\bibfnamefont {J.}~\bibnamefont {Behler}}, \ and\ \bibinfo {author} {\bibfnamefont {C.}~\bibnamefont {Dellago}},\ }\bibfield  {title} {\enquote {\bibinfo {title} {Parallel multistream training of high-dimensional neural network potentials},}\ }\href@noop {} {\bibfield  {journal} {\bibinfo  {journal} {Journal of Chemical Theory and Computation}\ }\textbf {\bibinfo {volume} {15}},\ \bibinfo {pages} {3075} (\bibinfo {year} {2019})}\BibitemShut {NoStop}%
\bibitem [{\citenamefont {Wang}\ \emph {et~al.}(2025)\citenamefont {Wang}, \citenamefont {Takaba}, \citenamefont {Chen}, \citenamefont {Wieder}, \citenamefont {Xu}, \citenamefont {Zhu}, \citenamefont {Zhang}, \citenamefont {Nagle}, \citenamefont {Yu}, \citenamefont {Wang}, \citenamefont {Cole}, \citenamefont {Rackers}, \citenamefont {Cho}, \citenamefont {Greener}, \citenamefont {Eastman}, \citenamefont {Martiniani},\ and\ \citenamefont {Tuckerman}}]{wang2025design}%
  \BibitemOpen
  \bibfield  {author} {\bibinfo {author} {\bibfnamefont {Y.}~\bibnamefont {Wang}}, \bibinfo {author} {\bibfnamefont {K.}~\bibnamefont {Takaba}}, \bibinfo {author} {\bibfnamefont {M.~S.}\ \bibnamefont {Chen}}, \bibinfo {author} {\bibfnamefont {M.}~\bibnamefont {Wieder}}, \bibinfo {author} {\bibfnamefont {Y.}~\bibnamefont {Xu}}, \bibinfo {author} {\bibfnamefont {T.}~\bibnamefont {Zhu}}, \bibinfo {author} {\bibfnamefont {J.~Z.~H.}\ \bibnamefont {Zhang}}, \bibinfo {author} {\bibfnamefont {A.}~\bibnamefont {Nagle}}, \bibinfo {author} {\bibfnamefont {K.}~\bibnamefont {Yu}}, \bibinfo {author} {\bibfnamefont {X.}~\bibnamefont {Wang}}, \bibinfo {author} {\bibfnamefont {D.~J.}\ \bibnamefont {Cole}}, \bibinfo {author} {\bibfnamefont {J.~A.}\ \bibnamefont {Rackers}}, \bibinfo {author} {\bibfnamefont {K.}~\bibnamefont {Cho}}, \bibinfo {author} {\bibfnamefont {J.~G.}\ \bibnamefont {Greener}}, \bibinfo {author} {\bibfnamefont {P.}~\bibnamefont {Eastman}}, \bibinfo {author} {\bibfnamefont {S.}~\bibnamefont {Martiniani}}, \
  and\ \bibinfo {author} {\bibfnamefont {M.~E.}\ \bibnamefont {Tuckerman}},\ }\bibfield  {title} {\enquote {\bibinfo {title} {On the design space between molecular mechanics and machine learning force fields},}\ }\href {\doibase 10.1063/5.0237876} {\bibfield  {journal} {\bibinfo  {journal} {Applied Physics Reviews}\ }\textbf {\bibinfo {volume} {12}},\ \bibinfo {pages} {021304} (\bibinfo {year} {2025})}\BibitemShut {NoStop}%
\bibitem [{\citenamefont {Batatia}\ \emph {et~al.}(2022)\citenamefont {Batatia}, \citenamefont {Kovacs}, \citenamefont {Simm}, \citenamefont {Ortner},\ and\ \citenamefont {Cs{\'a}nyi}}]{batatia2022mace}%
  \BibitemOpen
  \bibfield  {author} {\bibinfo {author} {\bibfnamefont {I.}~\bibnamefont {Batatia}}, \bibinfo {author} {\bibfnamefont {D.~P.}\ \bibnamefont {Kovacs}}, \bibinfo {author} {\bibfnamefont {G.}~\bibnamefont {Simm}}, \bibinfo {author} {\bibfnamefont {C.}~\bibnamefont {Ortner}}, \ and\ \bibinfo {author} {\bibfnamefont {G.}~\bibnamefont {Cs{\'a}nyi}},\ }\bibfield  {title} {\enquote {\bibinfo {title} {{MACE: Higher order equivariant message passing neural networks for fast and accurate force fields}},}\ }\href@noop {} {\bibfield  {journal} {\bibinfo  {journal} {Advances in Neural Information Processing Systems}\ }\textbf {\bibinfo {volume} {35}},\ \bibinfo {pages} {11423} (\bibinfo {year} {2022})}\BibitemShut {NoStop}%
\bibitem [{\citenamefont {Cheng}(2024)}]{cheng2024cartesian}%
  \BibitemOpen
  \bibfield  {author} {\bibinfo {author} {\bibfnamefont {B.}~\bibnamefont {Cheng}},\ }\bibfield  {title} {\enquote {\bibinfo {title} {Cartesian atomic cluster expansion for machine learning interatomic potentials},}\ }\href@noop {} {\bibfield  {journal} {\bibinfo  {journal} {npj Computational Materials}\ }\textbf {\bibinfo {volume} {10}},\ \bibinfo {pages} {157} (\bibinfo {year} {2024})}\BibitemShut {NoStop}%
\bibitem [{\citenamefont {Frenkel}\ and\ \citenamefont {Smit}(2023)}]{frenkel2023understanding}%
  \BibitemOpen
  \bibfield  {author} {\bibinfo {author} {\bibfnamefont {D.}~\bibnamefont {Frenkel}}\ and\ \bibinfo {author} {\bibfnamefont {B.}~\bibnamefont {Smit}},\ }\href {\doibase 10.1016/C2009-0-63921-0} {\emph {\bibinfo {title} {Understanding Molecular Simulation: From Algorithms to Applications}}},\ \bibinfo {edition} {3rd}\ ed.\ (\bibinfo  {publisher} {Academic Press},\ \bibinfo {year} {2023})\BibitemShut {NoStop}%
\bibitem [{\citenamefont {Kirkwood}\ and\ \citenamefont {Buff}(1951)}]{kirkwood1951statistical}%
  \BibitemOpen
  \bibfield  {author} {\bibinfo {author} {\bibfnamefont {J.~G.}\ \bibnamefont {Kirkwood}}\ and\ \bibinfo {author} {\bibfnamefont {F.~P.}\ \bibnamefont {Buff}},\ }\bibfield  {title} {\enquote {\bibinfo {title} {The statistical mechanical theory of solutions. i},}\ }\href@noop {} {\bibfield  {journal} {\bibinfo  {journal} {The Journal of Chemical Physics}\ }\textbf {\bibinfo {volume} {19}},\ \bibinfo {pages} {774} (\bibinfo {year} {1951})}\BibitemShut {NoStop}%
\bibitem [{\citenamefont {Ben-Naim}(2006)}]{ben2006molecular}%
  \BibitemOpen
  \bibfield  {author} {\bibinfo {author} {\bibfnamefont {A.}~\bibnamefont {Ben-Naim}},\ }\href {\doibase 10.1093/oso/9780199299690.001.0001} {\emph {\bibinfo {title} {Molecular Theory of Solutions}}}\ (\bibinfo  {publisher} {Oxford University Press},\ \bibinfo {year} {2006})\BibitemShut {NoStop}%
\bibitem [{\citenamefont {Mazzola}\ \emph {et~al.}(2018)\citenamefont {Mazzola}, \citenamefont {Helled},\ and\ \citenamefont {Sorella}}]{mazzola2018phase}%
  \BibitemOpen
  \bibfield  {author} {\bibinfo {author} {\bibfnamefont {G.}~\bibnamefont {Mazzola}}, \bibinfo {author} {\bibfnamefont {R.}~\bibnamefont {Helled}}, \ and\ \bibinfo {author} {\bibfnamefont {S.}~\bibnamefont {Sorella}},\ }\bibfield  {title} {\enquote {\bibinfo {title} {Phase diagram of hydrogen and a hydrogen-helium mixture at planetary conditions by quantum {Monte Carlo} simulations},}\ }\href@noop {} {\bibfield  {journal} {\bibinfo  {journal} {Physical Review Letters}\ }\textbf {\bibinfo {volume} {120}},\ \bibinfo {pages} {025701} (\bibinfo {year} {2018})}\BibitemShut {NoStop}%
\bibitem [{\citenamefont {Vorberger}\ \emph {et~al.}(2007)\citenamefont {Vorberger}, \citenamefont {Tamblyn}, \citenamefont {Militzer},\ and\ \citenamefont {Bonev}}]{Vorberger2007}%
  \BibitemOpen
  \bibfield  {author} {\bibinfo {author} {\bibfnamefont {J.}~\bibnamefont {Vorberger}}, \bibinfo {author} {\bibfnamefont {I.}~\bibnamefont {Tamblyn}}, \bibinfo {author} {\bibfnamefont {B.}~\bibnamefont {Militzer}}, \ and\ \bibinfo {author} {\bibfnamefont {S.~A.}\ \bibnamefont {Bonev}},\ }\bibfield  {title} {\enquote {\bibinfo {title} {Hydrogen-helium mixtures in the interiors of giant planets},}\ }\href {\doibase 10.1103/physrevb.75.024206} {\bibfield  {journal} {\bibinfo  {journal} {Physical Review B}\ }\textbf {\bibinfo {volume} {75}},\ \bibinfo {pages} {024206} (\bibinfo {year} {2007})}\BibitemShut {NoStop}%
\bibitem [{\citenamefont {Morales}\ \emph {et~al.}(2013{\natexlab{b}})\citenamefont {Morales}, \citenamefont {McMahon}, \citenamefont {Pierleoni},\ and\ \citenamefont {Ceperley}}]{morales2013nuclear}%
  \BibitemOpen
  \bibfield  {author} {\bibinfo {author} {\bibfnamefont {M.~A.}\ \bibnamefont {Morales}}, \bibinfo {author} {\bibfnamefont {J.~M.}\ \bibnamefont {McMahon}}, \bibinfo {author} {\bibfnamefont {C.}~\bibnamefont {Pierleoni}}, \ and\ \bibinfo {author} {\bibfnamefont {D.~M.}\ \bibnamefont {Ceperley}},\ }\bibfield  {title} {\enquote {\bibinfo {title} {Nuclear quantum effects and nonlocal exchange-correlation functionals applied to liquid hydrogen at high pressure},}\ }\href@noop {} {\bibfield  {journal} {\bibinfo  {journal} {Physical Review Letters}\ }\textbf {\bibinfo {volume} {110}},\ \bibinfo {pages} {065702} (\bibinfo {year} {2013}{\natexlab{b}})}\BibitemShut {NoStop}%
\bibitem [{\citenamefont {Lu}\ \emph {et~al.}(2019)\citenamefont {Lu}, \citenamefont {Kang}, \citenamefont {Wang}, \citenamefont {Gao},\ and\ \citenamefont {Dai}}]{lu2019towards}%
  \BibitemOpen
  \bibfield  {author} {\bibinfo {author} {\bibfnamefont {B.}~\bibnamefont {Lu}}, \bibinfo {author} {\bibfnamefont {D.}~\bibnamefont {Kang}}, \bibinfo {author} {\bibfnamefont {D.}~\bibnamefont {Wang}}, \bibinfo {author} {\bibfnamefont {T.}~\bibnamefont {Gao}}, \ and\ \bibinfo {author} {\bibfnamefont {J.}~\bibnamefont {Dai}},\ }\bibfield  {title} {\enquote {\bibinfo {title} {Towards the same line of liquid--liquid phase transition of dense hydrogen from various theoretical predictions},}\ }\href@noop {} {\bibfield  {journal} {\bibinfo  {journal} {Chinese Physics Letters}\ }\textbf {\bibinfo {volume} {36}},\ \bibinfo {pages} {103102} (\bibinfo {year} {2019})}\BibitemShut {NoStop}%
\bibitem [{\citenamefont {Hinz}\ \emph {et~al.}(2020)\citenamefont {Hinz}, \citenamefont {Karasiev}, \citenamefont {Hu}, \citenamefont {Zaghoo}, \citenamefont {Mej{\'\i}a-Rodr{\'\i}guez}, \citenamefont {Trickey},\ and\ \citenamefont {Calder{\'\i}n}}]{hinz2020fully}%
  \BibitemOpen
  \bibfield  {author} {\bibinfo {author} {\bibfnamefont {J.}~\bibnamefont {Hinz}}, \bibinfo {author} {\bibfnamefont {V.~V.}\ \bibnamefont {Karasiev}}, \bibinfo {author} {\bibfnamefont {S.~X.}\ \bibnamefont {Hu}}, \bibinfo {author} {\bibfnamefont {M.}~\bibnamefont {Zaghoo}}, \bibinfo {author} {\bibfnamefont {D.}~\bibnamefont {Mej{\'\i}a-Rodr{\'\i}guez}}, \bibinfo {author} {\bibfnamefont {S.~B.}\ \bibnamefont {Trickey}}, \ and\ \bibinfo {author} {\bibfnamefont {L.}~\bibnamefont {Calder{\'\i}n}},\ }\bibfield  {title} {\enquote {\bibinfo {title} {Fully consistent density functional theory determination of the insulator-metal transition boundary in warm dense hydrogen},}\ }\href@noop {} {\bibfield  {journal} {\bibinfo  {journal} {Physical Review Research}\ }\textbf {\bibinfo {volume} {2}},\ \bibinfo {pages} {032065} (\bibinfo {year} {2020})}\BibitemShut {NoStop}%
\bibitem [{\citenamefont {{van de Bund}}\ \emph {et~al.}(2021)\citenamefont {{van de Bund}}, \citenamefont {Wiebe},\ and\ \citenamefont {Ackland}}]{van2021isotope}%
  \BibitemOpen
  \bibfield  {author} {\bibinfo {author} {\bibfnamefont {S.}~\bibnamefont {{van de Bund}}}, \bibinfo {author} {\bibfnamefont {H.}~\bibnamefont {Wiebe}}, \ and\ \bibinfo {author} {\bibfnamefont {G.~J.}\ \bibnamefont {Ackland}},\ }\bibfield  {title} {\enquote {\bibinfo {title} {Isotope quantum effects in the metallization transition in liquid hydrogen},}\ }\href@noop {} {\bibfield  {journal} {\bibinfo  {journal} {Physical Review Letters}\ }\textbf {\bibinfo {volume} {126}},\ \bibinfo {pages} {225701} (\bibinfo {year} {2021})}\BibitemShut {NoStop}%
\bibitem [{\citenamefont {Bergermann}\ \emph {et~al.}(2024)\citenamefont {Bergermann}, \citenamefont {Kleindienst},\ and\ \citenamefont {Redmer}}]{bergermann2024nonmetal}%
  \BibitemOpen
  \bibfield  {author} {\bibinfo {author} {\bibfnamefont {A.}~\bibnamefont {Bergermann}}, \bibinfo {author} {\bibfnamefont {L.}~\bibnamefont {Kleindienst}}, \ and\ \bibinfo {author} {\bibfnamefont {R.}~\bibnamefont {Redmer}},\ }\bibfield  {title} {\enquote {\bibinfo {title} {{Nonmetal-to-metal transition in liquid hydrogen using density functional theory and the Heyd--Scuseria--Ernzerhof exchange-correlation functional}},}\ }\href {\doibase 10.1063/5.0241111} {\bibfield  {journal} {\bibinfo  {journal} {The Journal of Chemical Physics}\ }\textbf {\bibinfo {volume} {161}},\ \bibinfo {pages} {234303} (\bibinfo {year} {2024})}\BibitemShut {NoStop}%
\bibitem [{\citenamefont {Tenti}\ \emph {et~al.}(2025)\citenamefont {Tenti}, \citenamefont {J{\"a}ckl}, \citenamefont {Nakano}, \citenamefont {Rupp},\ and\ \citenamefont {Casula}}]{tenti2025hydrogen}%
  \BibitemOpen
  \bibfield  {author} {\bibinfo {author} {\bibfnamefont {G.}~\bibnamefont {Tenti}}, \bibinfo {author} {\bibfnamefont {B.}~\bibnamefont {J{\"a}ckl}}, \bibinfo {author} {\bibfnamefont {K.}~\bibnamefont {Nakano}}, \bibinfo {author} {\bibfnamefont {M.}~\bibnamefont {Rupp}}, \ and\ \bibinfo {author} {\bibfnamefont {M.}~\bibnamefont {Casula}},\ }\bibfield  {title} {\enquote {\bibinfo {title} {Hydrogen liquid-liquid transition from first principles and machine learning},}\ }\href@noop {} {\bibfield  {journal} {\bibinfo  {journal} {Physical Review B}\ }\textbf {\bibinfo {volume} {112}},\ \bibinfo {pages} {104208} (\bibinfo {year} {2025})}\BibitemShut {NoStop}%
\bibitem [{\citenamefont {Vazan}\ \emph {et~al.}(2016)\citenamefont {Vazan}, \citenamefont {Helled}, \citenamefont {Podolak},\ and\ \citenamefont {Kovetz}}]{vazan2016evolution}%
  \BibitemOpen
  \bibfield  {author} {\bibinfo {author} {\bibfnamefont {A.}~\bibnamefont {Vazan}}, \bibinfo {author} {\bibfnamefont {R.}~\bibnamefont {Helled}}, \bibinfo {author} {\bibfnamefont {M.}~\bibnamefont {Podolak}}, \ and\ \bibinfo {author} {\bibfnamefont {A.}~\bibnamefont {Kovetz}},\ }\bibfield  {title} {\enquote {\bibinfo {title} {{The evolution and internal structure of Jupiter and Saturn with compositional gradients}},}\ }\href@noop {} {\bibfield  {journal} {\bibinfo  {journal} {The Astrophysical Journal}\ }\textbf {\bibinfo {volume} {829}},\ \bibinfo {pages} {118} (\bibinfo {year} {2016})}\BibitemShut {NoStop}%
\bibitem [{\citenamefont {{Tejada Arevalo}}\ \emph {et~al.}(2025)\citenamefont {{Tejada Arevalo}}, \citenamefont {Sur}, \citenamefont {Su},\ and\ \citenamefont {Burrows}}]{arevalo2025jupiter}%
  \BibitemOpen
  \bibfield  {author} {\bibinfo {author} {\bibfnamefont {R.}~\bibnamefont {{Tejada Arevalo}}}, \bibinfo {author} {\bibfnamefont {A.}~\bibnamefont {Sur}}, \bibinfo {author} {\bibfnamefont {Y.}~\bibnamefont {Su}}, \ and\ \bibinfo {author} {\bibfnamefont {A.}~\bibnamefont {Burrows}},\ }\bibfield  {title} {\enquote {\bibinfo {title} {{Jupiter} evolutionary models incorporating stably stratified regions},}\ }\href {\doibase 10.3847/1538-4357/ada030} {\bibfield  {journal} {\bibinfo  {journal} {The Astrophysical Journal}\ }\textbf {\bibinfo {volume} {979}},\ \bibinfo {pages} {243} (\bibinfo {year} {2025})}\BibitemShut {NoStop}%
\bibitem [{\citenamefont {Sur}\ \emph {et~al.}(2025)\citenamefont {Sur}, \citenamefont {{Tejada Arevalo}}, \citenamefont {Su},\ and\ \citenamefont {Burrows}}]{sur2025simultaneous}%
  \BibitemOpen
  \bibfield  {author} {\bibinfo {author} {\bibfnamefont {A.}~\bibnamefont {Sur}}, \bibinfo {author} {\bibfnamefont {R.}~\bibnamefont {{Tejada Arevalo}}}, \bibinfo {author} {\bibfnamefont {Y.}~\bibnamefont {Su}}, \ and\ \bibinfo {author} {\bibfnamefont {A.}~\bibnamefont {Burrows}},\ }\bibfield  {title} {\enquote {\bibinfo {title} {Simultaneous evolutionary fits for {Jupiter and Saturn} incorporating fuzzy cores},}\ }\href {\doibase 10.3847/2041-8213/adad62} {\bibfield  {journal} {\bibinfo  {journal} {The Astrophysical Journal Letters}\ }\textbf {\bibinfo {volume} {980}},\ \bibinfo {pages} {L5} (\bibinfo {year} {2025})}\BibitemShut {NoStop}%
\bibitem [{\citenamefont {Bodenheimer}\ \emph {et~al.}(2025)\citenamefont {Bodenheimer}, \citenamefont {Stevenson}, \citenamefont {Lissauer},\ and\ \citenamefont {D'Angelo}}]{bodenheimer2025formation}%
  \BibitemOpen
  \bibfield  {author} {\bibinfo {author} {\bibfnamefont {P.}~\bibnamefont {Bodenheimer}}, \bibinfo {author} {\bibfnamefont {D.~J.}\ \bibnamefont {Stevenson}}, \bibinfo {author} {\bibfnamefont {J.~J.}\ \bibnamefont {Lissauer}}, \ and\ \bibinfo {author} {\bibfnamefont {G.}~\bibnamefont {D'Angelo}},\ }\bibfield  {title} {\enquote {\bibinfo {title} {{Formation and evolution simulations of Saturn, including composition gradients and helium immiscibility}},}\ }\href {\doibase 10.3847/PSJ/add0b0} {\bibfield  {journal} {\bibinfo  {journal} {The Planetary Science Journal}\ }\textbf {\bibinfo {volume} {6}},\ \bibinfo {pages} {143} (\bibinfo {year} {2025})}\BibitemShut {NoStop}%
\bibitem [{\citenamefont {Markham}\ and\ \citenamefont {Guillot}(2024)}]{markham2024stable}%
  \BibitemOpen
  \bibfield  {author} {\bibinfo {author} {\bibfnamefont {S.}~\bibnamefont {Markham}}\ and\ \bibinfo {author} {\bibfnamefont {T.}~\bibnamefont {Guillot}},\ }\bibfield  {title} {\enquote {\bibinfo {title} {{Stable stratification of the helium rain layer yields vastly different interiors and magnetic fields for Jupiter and Saturn}},}\ }\href@noop {} {\bibfield  {journal} {\bibinfo  {journal} {The Planetary Science Journal}\ }\textbf {\bibinfo {volume} {5}},\ \bibinfo {pages} {269} (\bibinfo {year} {2024})}\BibitemShut {NoStop}%
\bibitem [{\citenamefont {Sur}\ \emph {et~al.}(2026)\citenamefont {Sur}, \citenamefont {{Tejada Arevalo}}, \citenamefont {Burrows},\ and\ \citenamefont {Chen}}]{sur2026next}%
  \BibitemOpen
  \bibfield  {author} {\bibinfo {author} {\bibfnamefont {A.}~\bibnamefont {Sur}}, \bibinfo {author} {\bibfnamefont {R.}~\bibnamefont {{Tejada Arevalo}}}, \bibinfo {author} {\bibfnamefont {A.}~\bibnamefont {Burrows}}, \ and\ \bibinfo {author} {\bibfnamefont {Y.-X.}\ \bibnamefont {Chen}},\ }\bibfield  {title} {\enquote {\bibinfo {title} {Next-generation improvements in giant-exoplanet evolutionary and structural models},}\ }\href {\doibase 10.3847/1538-4357/ae3a85} {\bibfield  {journal} {\bibinfo  {journal} {The Astrophysical Journal}\ }\textbf {\bibinfo {volume} {998}},\ \bibinfo {pages} {305} (\bibinfo {year} {2026})}\BibitemShut {NoStop}%
\bibitem [{\citenamefont {Fortney}\ \emph {et~al.}(2011)\citenamefont {Fortney}, \citenamefont {Ikoma}, \citenamefont {Nettelmann}, \citenamefont {Guillot},\ and\ \citenamefont {Marley}}]{fortney2011self}%
  \BibitemOpen
  \bibfield  {author} {\bibinfo {author} {\bibfnamefont {J.~J.}\ \bibnamefont {Fortney}}, \bibinfo {author} {\bibfnamefont {M.}~\bibnamefont {Ikoma}}, \bibinfo {author} {\bibfnamefont {N.}~\bibnamefont {Nettelmann}}, \bibinfo {author} {\bibfnamefont {T.}~\bibnamefont {Guillot}}, \ and\ \bibinfo {author} {\bibfnamefont {M.}~\bibnamefont {Marley}},\ }\bibfield  {title} {\enquote {\bibinfo {title} {Self-consistent model atmospheres and the cooling of the solar system's giant planets},}\ }\href@noop {} {\bibfield  {journal} {\bibinfo  {journal} {The Astrophysical Journal}\ }\textbf {\bibinfo {volume} {729}},\ \bibinfo {pages} {32} (\bibinfo {year} {2011})}\BibitemShut {NoStop}%
\bibitem [{\citenamefont {P{\"u}stow}\ \emph {et~al.}(2016)\citenamefont {P{\"u}stow}, \citenamefont {Nettelmann}, \citenamefont {Lorenzen},\ and\ \citenamefont {Redmer}}]{pustow2016h}%
  \BibitemOpen
  \bibfield  {author} {\bibinfo {author} {\bibfnamefont {R.}~\bibnamefont {P{\"u}stow}}, \bibinfo {author} {\bibfnamefont {N.}~\bibnamefont {Nettelmann}}, \bibinfo {author} {\bibfnamefont {W.}~\bibnamefont {Lorenzen}}, \ and\ \bibinfo {author} {\bibfnamefont {R.}~\bibnamefont {Redmer}},\ }\bibfield  {title} {\enquote {\bibinfo {title} {{H/He demixing and the cooling behavior of Saturn}},}\ }\href@noop {} {\bibfield  {journal} {\bibinfo  {journal} {Icarus}\ }\textbf {\bibinfo {volume} {267}},\ \bibinfo {pages} {323} (\bibinfo {year} {2016})}\BibitemShut {NoStop}%
\bibitem [{\citenamefont {Preising}\ \emph {et~al.}(2023)\citenamefont {Preising}, \citenamefont {French}, \citenamefont {Mankovich}, \citenamefont {Soubiran},\ and\ \citenamefont {Redmer}}]{preising2023material}%
  \BibitemOpen
  \bibfield  {author} {\bibinfo {author} {\bibfnamefont {M.}~\bibnamefont {Preising}}, \bibinfo {author} {\bibfnamefont {M.}~\bibnamefont {French}}, \bibinfo {author} {\bibfnamefont {C.}~\bibnamefont {Mankovich}}, \bibinfo {author} {\bibfnamefont {F.}~\bibnamefont {Soubiran}}, \ and\ \bibinfo {author} {\bibfnamefont {R.}~\bibnamefont {Redmer}},\ }\bibfield  {title} {\enquote {\bibinfo {title} {Material properties of {Saturn’s} interior from {\it ab initio} simulations},}\ }\href@noop {} {\bibfield  {journal} {\bibinfo  {journal} {The Astrophysical Journal Supplement Series}\ }\textbf {\bibinfo {volume} {269}},\ \bibinfo {pages} {47} (\bibinfo {year} {2023})}\BibitemShut {NoStop}%
\bibitem [{\citenamefont {Mankovich}\ and\ \citenamefont {Fuller}(2021)}]{mankovich2021diffuse}%
  \BibitemOpen
  \bibfield  {author} {\bibinfo {author} {\bibfnamefont {C.~R.}\ \bibnamefont {Mankovich}}\ and\ \bibinfo {author} {\bibfnamefont {J.}~\bibnamefont {Fuller}},\ }\bibfield  {title} {\enquote {\bibinfo {title} {{A diffuse core in Saturn revealed by ring seismology}},}\ }\href@noop {} {\bibfield  {journal} {\bibinfo  {journal} {Nature Astronomy}\ }\textbf {\bibinfo {volume} {5}},\ \bibinfo {pages} {1103} (\bibinfo {year} {2021})}\BibitemShut {NoStop}%
\bibitem [{\citenamefont {M{\"u}ller}\ and\ \citenamefont {Helled}(2023)}]{muller2023warm}%
  \BibitemOpen
  \bibfield  {author} {\bibinfo {author} {\bibfnamefont {S.}~\bibnamefont {M{\"u}ller}}\ and\ \bibinfo {author} {\bibfnamefont {R.}~\bibnamefont {Helled}},\ }\bibfield  {title} {\enquote {\bibinfo {title} {Warm giant exoplanet characterisation: {Current state}, challenges and outlook},}\ }\href@noop {} {\bibfield  {journal} {\bibinfo  {journal} {Frontiers in Astronomy and Space Sciences}\ }\textbf {\bibinfo {volume} {10}},\ \bibinfo {pages} {1179000} (\bibinfo {year} {2023})}\BibitemShut {NoStop}%
\bibitem [{\citenamefont {Kresse}\ and\ \citenamefont {Furthm{\"u}ller}(1996)}]{kresse1996efficient}%
  \BibitemOpen
  \bibfield  {author} {\bibinfo {author} {\bibfnamefont {G.}~\bibnamefont {Kresse}}\ and\ \bibinfo {author} {\bibfnamefont {J.}~\bibnamefont {Furthm{\"u}ller}},\ }\bibfield  {title} {\enquote {\bibinfo {title} {Efficient iterative schemes for {\it ab initio} total-energy calculations using a plane-wave basis set},}\ }\href {\doibase 10.1103/PhysRevB.54.11169} {\bibfield  {journal} {\bibinfo  {journal} {Physical Review B}\ }\textbf {\bibinfo {volume} {54}},\ \bibinfo {pages} {11169} (\bibinfo {year} {1996})}\BibitemShut {NoStop}%
\bibitem [{\citenamefont {Behler}\ and\ \citenamefont {Parrinello}(2007)}]{behler2007generalized}%
  \BibitemOpen
  \bibfield  {author} {\bibinfo {author} {\bibfnamefont {J.}~\bibnamefont {Behler}}\ and\ \bibinfo {author} {\bibfnamefont {M.}~\bibnamefont {Parrinello}},\ }\bibfield  {title} {\enquote {\bibinfo {title} {Generalized neural-network representation of high-dimensional potential-energy surfaces},}\ }\href@noop {} {\bibfield  {journal} {\bibinfo  {journal} {Physical Review Letters}\ }\textbf {\bibinfo {volume} {98}},\ \bibinfo {pages} {146401} (\bibinfo {year} {2007})}\BibitemShut {NoStop}%
\bibitem [{\citenamefont {Thompson}\ \emph {et~al.}(2022)\citenamefont {Thompson}, \citenamefont {Aktulga}, \citenamefont {Berger}, \citenamefont {Bolintineanu}, \citenamefont {Brown}, \citenamefont {Crozier}, \citenamefont {{in 't Veld}}, \citenamefont {Kohlmeyer}, \citenamefont {Moore}, \citenamefont {Nguyen}, \citenamefont {Shan}, \citenamefont {Stevens}, \citenamefont {Tranchida}, \citenamefont {Trott},\ and\ \citenamefont {Plimpton}}]{LAMMPS}%
  \BibitemOpen
  \bibfield  {author} {\bibinfo {author} {\bibfnamefont {A.~P.}\ \bibnamefont {Thompson}}, \bibinfo {author} {\bibfnamefont {H.~M.}\ \bibnamefont {Aktulga}}, \bibinfo {author} {\bibfnamefont {R.}~\bibnamefont {Berger}}, \bibinfo {author} {\bibfnamefont {D.~S.}\ \bibnamefont {Bolintineanu}}, \bibinfo {author} {\bibfnamefont {W.~M.}\ \bibnamefont {Brown}}, \bibinfo {author} {\bibfnamefont {P.~S.}\ \bibnamefont {Crozier}}, \bibinfo {author} {\bibfnamefont {P.~J.}\ \bibnamefont {{in 't Veld}}}, \bibinfo {author} {\bibfnamefont {A.}~\bibnamefont {Kohlmeyer}}, \bibinfo {author} {\bibfnamefont {S.~G.}\ \bibnamefont {Moore}}, \bibinfo {author} {\bibfnamefont {T.~D.}\ \bibnamefont {Nguyen}}, \bibinfo {author} {\bibfnamefont {R.}~\bibnamefont {Shan}}, \bibinfo {author} {\bibfnamefont {M.~J.}\ \bibnamefont {Stevens}}, \bibinfo {author} {\bibfnamefont {J.}~\bibnamefont {Tranchida}}, \bibinfo {author} {\bibfnamefont {C.}~\bibnamefont {Trott}}, \ and\ \bibinfo {author} {\bibfnamefont {S.~J.}\ \bibnamefont {Plimpton}},\
  }\bibfield  {title} {\enquote {\bibinfo {title} {{LAMMPS} - a flexible simulation tool for particle-based materials modeling at the atomic, meso, and continuum scales},}\ }\href {\doibase 10.1016/j.cpc.2021.108171} {\bibfield  {journal} {\bibinfo  {journal} {Computer Physics Communications}\ }\textbf {\bibinfo {volume} {271}},\ \bibinfo {pages} {108171} (\bibinfo {year} {2022})}\BibitemShut {NoStop}%
\bibitem [{\citenamefont {Larsen}\ \emph {et~al.}(2017)\citenamefont {Larsen}, \citenamefont {Mortensen}, \citenamefont {Blomqvist}, \citenamefont {Castelli}, \citenamefont {Christensen}, \citenamefont {Du{\l}ak}, \citenamefont {Friis}, \citenamefont {Groves}, \citenamefont {Hammer}, \citenamefont {Hargus}, \citenamefont {Hermes}, \citenamefont {Jennings}, \citenamefont {Jensen}, \citenamefont {Kermode}, \citenamefont {Kitchin}, \citenamefont {Kolsbjerg}, \citenamefont {Kubal}, \citenamefont {Kaasbjerg}, \citenamefont {Lysgaard}, \citenamefont {Maronsson}, \citenamefont {Maxson}, \citenamefont {Olsen}, \citenamefont {Pastewka}, \citenamefont {Peterson}, \citenamefont {Rostgaard}, \citenamefont {Schi{\o}tz}, \citenamefont {Sch{\"u}tt}, \citenamefont {Strange}, \citenamefont {Thygesen}, \citenamefont {Vegge}, \citenamefont {Vilhelmsen}, \citenamefont {Walter}, \citenamefont {Zeng},\ and\ \citenamefont {Jacobsen}}]{larsen2017atomic}%
  \BibitemOpen
  \bibfield  {author} {\bibinfo {author} {\bibfnamefont {A.~H.}\ \bibnamefont {Larsen}}, \bibinfo {author} {\bibfnamefont {J.~J.}\ \bibnamefont {Mortensen}}, \bibinfo {author} {\bibfnamefont {J.}~\bibnamefont {Blomqvist}}, \bibinfo {author} {\bibfnamefont {I.~E.}\ \bibnamefont {Castelli}}, \bibinfo {author} {\bibfnamefont {R.}~\bibnamefont {Christensen}}, \bibinfo {author} {\bibfnamefont {M.}~\bibnamefont {Du{\l}ak}}, \bibinfo {author} {\bibfnamefont {J.}~\bibnamefont {Friis}}, \bibinfo {author} {\bibfnamefont {M.~N.}\ \bibnamefont {Groves}}, \bibinfo {author} {\bibfnamefont {B.}~\bibnamefont {Hammer}}, \bibinfo {author} {\bibfnamefont {C.}~\bibnamefont {Hargus}}, \bibinfo {author} {\bibfnamefont {E.~D.}\ \bibnamefont {Hermes}}, \bibinfo {author} {\bibfnamefont {P.~C.}\ \bibnamefont {Jennings}}, \bibinfo {author} {\bibfnamefont {P.~B.}\ \bibnamefont {Jensen}}, \bibinfo {author} {\bibfnamefont {J.}~\bibnamefont {Kermode}}, \bibinfo {author} {\bibfnamefont {J.~R.}\ \bibnamefont {Kitchin}}, \bibinfo {author}
  {\bibfnamefont {E.~L.}\ \bibnamefont {Kolsbjerg}}, \bibinfo {author} {\bibfnamefont {J.}~\bibnamefont {Kubal}}, \bibinfo {author} {\bibfnamefont {K.}~\bibnamefont {Kaasbjerg}}, \bibinfo {author} {\bibfnamefont {S.}~\bibnamefont {Lysgaard}}, \bibinfo {author} {\bibfnamefont {J.~B.}\ \bibnamefont {Maronsson}}, \bibinfo {author} {\bibfnamefont {T.}~\bibnamefont {Maxson}}, \bibinfo {author} {\bibfnamefont {T.}~\bibnamefont {Olsen}}, \bibinfo {author} {\bibfnamefont {L.}~\bibnamefont {Pastewka}}, \bibinfo {author} {\bibfnamefont {A.}~\bibnamefont {Peterson}}, \bibinfo {author} {\bibfnamefont {C.}~\bibnamefont {Rostgaard}}, \bibinfo {author} {\bibfnamefont {J.}~\bibnamefont {Schi{\o}tz}}, \bibinfo {author} {\bibfnamefont {O.}~\bibnamefont {Sch{\"u}tt}}, \bibinfo {author} {\bibfnamefont {M.}~\bibnamefont {Strange}}, \bibinfo {author} {\bibfnamefont {K.~S.}\ \bibnamefont {Thygesen}}, \bibinfo {author} {\bibfnamefont {T.}~\bibnamefont {Vegge}}, \bibinfo {author} {\bibfnamefont {L.}~\bibnamefont {Vilhelmsen}},
  \bibinfo {author} {\bibfnamefont {M.}~\bibnamefont {Walter}}, \bibinfo {author} {\bibfnamefont {Z.}~\bibnamefont {Zeng}}, \ and\ \bibinfo {author} {\bibfnamefont {K.~W.}\ \bibnamefont {Jacobsen}},\ }\bibfield  {title} {\enquote {\bibinfo {title} {The atomic simulation environment—a {Python} library for working with atoms},}\ }\href {\doibase 10.1088/1361-648X/aa680e} {\bibfield  {journal} {\bibinfo  {journal} {Journal of Physics: Condensed Matter}\ }\textbf {\bibinfo {volume} {29}},\ \bibinfo {pages} {273002} (\bibinfo {year} {2017})}\BibitemShut {NoStop}%
\bibitem [{\citenamefont {Tuckerman}(2023)}]{tuckerman2023statistical}%
  \BibitemOpen
  \bibfield  {author} {\bibinfo {author} {\bibfnamefont {M.~E.}\ \bibnamefont {Tuckerman}},\ }\href {\doibase 10.1093/oso/9780198825562.001.0001} {\emph {\bibinfo {title} {Statistical Mechanics: Theory and Molecular Simulation}}},\ \bibinfo {edition} {2nd}\ ed.\ (\bibinfo  {publisher} {Oxford University Press},\ \bibinfo {year} {2023})\BibitemShut {NoStop}%
\bibitem [{\citenamefont {Ceriotti}\ \emph {et~al.}(2014)\citenamefont {Ceriotti}, \citenamefont {More},\ and\ \citenamefont {Manolopoulos}}]{ceriotti2014pi}%
  \BibitemOpen
  \bibfield  {author} {\bibinfo {author} {\bibfnamefont {M.}~\bibnamefont {Ceriotti}}, \bibinfo {author} {\bibfnamefont {J.}~\bibnamefont {More}}, \ and\ \bibinfo {author} {\bibfnamefont {D.~E.}\ \bibnamefont {Manolopoulos}},\ }\bibfield  {title} {\enquote {\bibinfo {title} {{i-PI: A Python interface for ab initio path integral molecular dynamics simulations}},}\ }\href@noop {} {\bibfield  {journal} {\bibinfo  {journal} {Computer Physics Communications}\ }\textbf {\bibinfo {volume} {185}},\ \bibinfo {pages} {1019} (\bibinfo {year} {2014})}\BibitemShut {NoStop}%
\bibitem [{\citenamefont {Litman}\ \emph {et~al.}(2024)\citenamefont {Litman}, \citenamefont {Kapil}, \citenamefont {Feldman}, \citenamefont {Tisi}, \citenamefont {Begu{\v{s}}i{\'c}}, \citenamefont {Fidanyan}, \citenamefont {Fraux}, \citenamefont {Higer}, \citenamefont {Kellner}, \citenamefont {Li}, \citenamefont {P{\'o}s}, \citenamefont {Stocco}, \citenamefont {Trenins}, \citenamefont {Hirshberg}, \citenamefont {Rossi},\ and\ \citenamefont {Ceriotti}}]{litman2024pi}%
  \BibitemOpen
  \bibfield  {author} {\bibinfo {author} {\bibfnamefont {Y.}~\bibnamefont {Litman}}, \bibinfo {author} {\bibfnamefont {V.}~\bibnamefont {Kapil}}, \bibinfo {author} {\bibfnamefont {Y.~M.~Y.}\ \bibnamefont {Feldman}}, \bibinfo {author} {\bibfnamefont {D.}~\bibnamefont {Tisi}}, \bibinfo {author} {\bibfnamefont {T.}~\bibnamefont {Begu{\v{s}}i{\'c}}}, \bibinfo {author} {\bibfnamefont {K.}~\bibnamefont {Fidanyan}}, \bibinfo {author} {\bibfnamefont {G.}~\bibnamefont {Fraux}}, \bibinfo {author} {\bibfnamefont {J.}~\bibnamefont {Higer}}, \bibinfo {author} {\bibfnamefont {M.}~\bibnamefont {Kellner}}, \bibinfo {author} {\bibfnamefont {T.~E.}\ \bibnamefont {Li}}, \bibinfo {author} {\bibfnamefont {E.~S.}\ \bibnamefont {P{\'o}s}}, \bibinfo {author} {\bibfnamefont {E.}~\bibnamefont {Stocco}}, \bibinfo {author} {\bibfnamefont {G.}~\bibnamefont {Trenins}}, \bibinfo {author} {\bibfnamefont {B.}~\bibnamefont {Hirshberg}}, \bibinfo {author} {\bibfnamefont {M.}~\bibnamefont {Rossi}}, \ and\ \bibinfo {author} {\bibfnamefont
  {M.}~\bibnamefont {Ceriotti}},\ }\bibfield  {title} {\enquote {\bibinfo {title} {{{i-PI} 3.0}: A flexible and efficient framework for advanced atomistic simulations},}\ }\href {\doibase 10.1063/5.0215869} {\bibfield  {journal} {\bibinfo  {journal} {The Journal of Chemical Physics}\ }\textbf {\bibinfo {volume} {161}},\ \bibinfo {pages} {062504} (\bibinfo {year} {2024})}\BibitemShut {NoStop}%
\bibitem [{\citenamefont {Ceriotti}\ \emph {et~al.}(2010)\citenamefont {Ceriotti}, \citenamefont {Parrinello}, \citenamefont {Markland},\ and\ \citenamefont {Manolopoulos}}]{ceriotti2010efficient}%
  \BibitemOpen
  \bibfield  {author} {\bibinfo {author} {\bibfnamefont {M.}~\bibnamefont {Ceriotti}}, \bibinfo {author} {\bibfnamefont {M.}~\bibnamefont {Parrinello}}, \bibinfo {author} {\bibfnamefont {T.~E.}\ \bibnamefont {Markland}}, \ and\ \bibinfo {author} {\bibfnamefont {D.~E.}\ \bibnamefont {Manolopoulos}},\ }\bibfield  {title} {\enquote {\bibinfo {title} {Efficient stochastic thermostatting of path integral molecular dynamics},}\ }\href {\doibase 10.1063/1.3489925} {\bibfield  {journal} {\bibinfo  {journal} {The Journal of Chemical Physics}\ }\textbf {\bibinfo {volume} {133}},\ \bibinfo {pages} {124104} (\bibinfo {year} {2010})}\BibitemShut {NoStop}%
\bibitem [{\citenamefont {Wang}\ \emph {et~al.}(2026)\citenamefont {Wang}, \citenamefont {Hamel},\ and\ \citenamefont {Cheng}}]{wang_2026_21632034}%
  \BibitemOpen
  \bibfield  {author} {\bibinfo {author} {\bibfnamefont {X.}~\bibnamefont {Wang}}, \bibinfo {author} {\bibfnamefont {S.}~\bibnamefont {Hamel}}, \ and\ \bibinfo {author} {\bibfnamefont {B.}~\bibnamefont {Cheng}},\ }\href {\doibase 10.5281/zenodo.21632034} {\enquote {\bibinfo {title} {Source data for high pressure {H-He} mixtures},}\ } (\bibinfo {year} {2026})\BibitemShut {NoStop}%
\end{thebibliography}
\end{document}